\documentclass[11pt]{article}
\usepackage{graphicx}
\usepackage{amssymb}
\usepackage{epstopdf}
\usepackage[small]{caption}
\def\R{{\mathbb R}}
\def\O{{\mathcal O}}
\def\diff{\mathrm{Diff}}
\def\eps{\varepsilon}

\def\eqalign#1{\null\,\vcenter{\openup\jot \mathsurround=0pt \ialign{\strut
     \hfil$\displaystyle{##}$&$ \displaystyle{{}##}$\hfil \crcr#1\crcr}}\,}

\title{$N$-Particle dynamics of the Euler equations for planar diffeomorphisms}
\author{Robert I McLachlan$^1$ and Stephen Marsland$^2$\\
$^1$ IFS, Massey University, Palmerston North, New Zealand\\
$^2$ IIS\&T, Massey University, Palmerston North, New Zealand\\
Email: \{r.mclachlan,s.r.marsland\}@massey.ac.nz}
\date{}

\begin{document}
\maketitle

\abstract{The Euler equations associated with diffeomorphism groups have received much recent study because
of their links with fluid dynamics, computer vision, and mechanics. In this paper, we consider the dynamics of $N$ point particles or `blobs' moving
under the action of the Euler equations associated with the group of diffeomorphisms of the plane in a variety of different metrics. The 2 body problem is always integrable, and we analyze its phase portrait under different metrics. In particular, we show that 2-body capturing orbits (in which the distances between the particles tend to 0 as $t \to \infty$) can occur when the kernel is sufficiently smooth and the relative initial velocity of the particles is sufficiently large. We compute the dynamics of these `dipoles' with respect to other test particles, and supplement the calculations with simulations for larger $N$ that illustrate the different regimes.}

\section{Introduction}

The Euler equations for planar diffeomorphisms are
\begin{equation}
\dot{m} + {u} \cdot \nabla {m} + \nabla {u}^T \cdot {m} + {m} ( \mathrm{div} \:  {u} ) = 0, 
\label{Euler1}
\end{equation}
where $\dot{m}$ denotes differentiation with respect to time, $u(x,t)$ ($u,x\in\R^n$, $t\in \R$) is a velocity field, and $m(x,t)$ its associated momentum. The velocity $u$ and momentum $m$ are related by 
\begin{equation}
\label{Euler2}
m = \mathcal{A} u
\end{equation}
where $\mathcal{A}$ is an elliptic operator (e.g.  $\mathcal{A} = (1-\nabla^2)^k$) called the inertia operator.

There has been much recent interest in the Euler equations (\ref{Euler1}) because they arise in several different fields, including computer vision
and fluid dynamics. In computer vision they appear in two and three dimensions in the field of image registration, such as in the averaged template matching equations~\cite{MillerTrouveYounes02,Mumford98} and the
geodesic interpolating clamped-plate spline~\cite{MarslandTwining04}, while in fluid dynamics they are the limiting case of the shallow water wave Camassa-Holm equations in one and two dimensions~\cite{HolmMarsden04}, and the motion of an ideal incompressible fluid corresponds to geodesics on the group of volume-preserving diffeomorphisms~\cite{Arnold66,EbinMarsden70}

Euler equations such as Eq. (\ref{Euler1}) have a natural geometric origin, which we sketch here although these details are not needed in the paper. Let $\mathfrak{G}$ be a Lie or diffeomorphism group  with  Lie algebra $\mathfrak{g}$.
Let $\mathfrak{G}$ be equipped with a left- or right-invariant metric that restricts to a metric
$\langle\!\langle , \rangle\!\rangle$ 
on $\mathfrak{g}$. Typically, this metric is defined by a linear {\it inertia operator} $\mathcal{A} \colon \mathfrak{g}\to \mathfrak{g}^*$ via 
\begin{equation}\langle\!\langle u,v \rangle\!\rangle := \langle u, \mathcal{A}v \rangle.\end{equation}
The geodesic equation on $T\mathfrak{G}$ can be reduced to give a non-canonical Hamiltonian evolution equation on $\mathfrak{g}$ called the {\it Euler--Poincar\'e} equation, or transferred (via the Legendre transform, $\mathfrak{g} \to \mathfrak{g}^*$, which in this case is $u \to m := \mathcal{A} u$) to a Lie-Poisson system on $\mathfrak{g}^*$ called the {\it Euler} equations 
\begin{equation}
\dot{m} = \pm \mathrm{ad}^*_{\mathcal{A}^{-1} m} m,
\label{eq:Euler}
\end{equation}
where the sign is $+$ for left- and $-$ for right-invariant metrics. The variable $m\in\mathfrak{g}^*$ is called the {\it momentum} and $u\in\mathfrak{g}$ the {\it velocity}.

The  most famous Euler equation on a Lie group is the equation of motion of a free rigid body ($\mathfrak{G}=SO(3)$, $m=$ body angular momentum). An infinite-dimensional example is the Landau-Lifshitz equation on the loop group $\mathfrak{G}=C^{\infty}(\mathbb{R},SO(3))$. Examples on diffeomorphism groups include the Euler fluid equations on the group $\mathfrak{G}=\diff_{\rm vol}(\R^n)$ of volume-preserving diffeomorphisms with respect to the $L^2$ metric ($m=$ vorticity)~\cite{EbinMarsden70}, the Camassa-Holm equation ($\mathfrak{G}=\diff(\mathbb{S}^1)$, $H^1$ metric)~\cite{HolmMarsden04}, and the second-grade fluid equations ($\mathfrak{G}=\diff_{\rm vol}(\R^n)$, $H^1$ metric)~\cite{OliverShkoller01}.

There has been recent interest in the use of diffeomorphic deformations (`warps') to align groups of medical images. This alignment allows the range of variation in the appearance of biological structures to be measured, providing an approach to automated disease diagnosis. One approach to this problem, known as Computational Anatomy, is to introduce group actions as deformable templates that are warped via the actions of a group onto other images~\cite{Grenander98}. This work has been fundamental to a large amount of research on aligning images through landmark matching, where corresponding points are defined on a set of images, and diffeomorphic warps used to align them. The corresponding group for image analysis is the full diffeomorphism group, not the volume-preserving subgroup. Under a right-invariant Riemannian metric, it can be shown that the geodesics of the motion of a set of landmarks can be computed as an optimisation problem. See \cite{MarslandTwining04,MillerTrouveYounes02} for an overview. 

For $\mathfrak{G}=\diff(\R^n)$ the Euler equations are given by Eqs. (\ref{Euler1},\ref{Euler2}) (see~\cite{HolmMarsdenRatiu98,Mumford98} for further details). The inverse of the inertia operator $\mathcal{A}$ is given by convolution with the Green's function $\mathbf{G}$ of $\mathcal{A}$, i.e., $u = \mathbf{G}*m$, where $*$ denotes convolution and $\mathcal{A}\mathbf{G}(x,x') = \delta(x-x')$ for $x,x'\in\R^n$. We shall only consider rotationally invariant and diagonal $\mathcal{A}$; in this case $\mathbf{G}(x,x') = G(\|x-x'\|)$ for a scalar function $G$, which we call the {\em kernel} of $\mathbf{G}$. Examples in the literature include Gaussian \cite{HolmRatnanatherTrouveYounes04} and Bessel function kernels \cite{OliverShkoller01} associated with the $H^1$ metric. In this paper, we consider a family of kernels associated with $H^k$ metrics for various $k$, which includes the Gaussian kernel in the limit as $k \to \infty$.

A striking feature of Euler equations on diffeomorphism groups is that they admit (formally, at least) exact solutions in which the momentum is concentrated at a finite set of points that we call particles. For fluid equations these are point vortices, which are widely studied both in their own right and as a means of approximating the evolution of smooth or other vorticities. At first sight it is remarkable that a PDE should have such a finite-dimensional reduction, unrelated to integrability. (Such a reduction, however, is not unique to Euler equations. On the other hand, there are many particle-based numerical methods, collectively known as Smoothed Particle Hydrodynamics~\cite{Monaghan92}, which are not exact solutions of the PDE they approximate.) 

For the 2D and 3D Euler fluid equations, convergence of the point vortex solutions to solutions for smooth initial data has been established \cite{BealeMajda82}. The speed of convergence can be improved by smoothing out the point vortices to {\it vortex blobs} \cite{Chorin73}, even though the (e.g., Gaussian) blobs are no longer an exact solution of the Euler equations. Instead, their evolution can be regarded as that of delta-functions under a slightly different inertia operator. Altering the shape of the blob (which corresponds to the kernel) is equivalent to altering the metric. In this way we are lead to consider a wide class of metrics. For an introduction to many-body problems, particularly arising in Newtonian and relativistic mechanics, see~\cite{Calogero01}. Point vortices defined over Dirac delta-functions, and the regularised version, vortex blob methods, are introduced in~\cite{Chorin73}; see~\cite{Leonard80} for a review of their use.  

The methods used for image registration and template matching of images in~\cite{MillerTrouveYounes02,MarslandTwining04}, and references therein, correspond to computing the motion of particles in $\diff(\mathbb{R}^2)$. Various metrics have been used, including Gaussians, $H^1$, and $H^k$. Diffeomorphic image registration can be considered as a boundary value problem, since the start and end points of the particles are defined by the images under consideration. However, it can equivalently be treated as an initial value problem under the definition of the starting point and initial velocity of the particles~\cite{Vaillant04,MarslandTwining04}. This paper provides an initial study of the dynamics of particles for the group $\diff(\R^2)$ of all diffeomorphisms of the plane, as are used in the image registration framework. Following the approach taken in fluid dynamics, we consider a set of $N$ particles moving under the action of a kernel $G$. 


We begin the paper  by explicitly computing the Hamiltonian for the case of $N$ particles, and computing the equations of motion in terms of the Green's function. Following this, in section~\ref{findkernel} we use delta-function particles and Helmholtz-style inertia operators to find the kernels $G_k$ corresponding to $H^k$ metrics. The limit of these, $H^{\infty}$, corresponds to a Gaussian kernel for delta-function particles, where the inertia operator tends to $\exp (- \eps ^2 \nabla^2)$.  We show how even when blobs are required, rather than particles, we can still consider delta-function particles through a change in the inertia operator. 

We then shift our attention to the case $N=2$, and examine the simplest possible interaction of particles. This two body problem is always integrable. We derive the reduced Hamiltonian and consider how the particle dynamics change for different  metrics. The most striking feature of the dynamics is the phenomenon of {\it particle capture}, in which particles cohere or stick together.  In a typical capture orbit, the distance between two particles is $\mathcal{O}(e^{-at})$ as $t\to\infty$, but there is no actual collision. We identify the regimes of capture and (more traditional) scattering orbits. For different metrics, we also show the phase portraits of the dynamics, and the appearance of typical scattering orbits. The analysis shows that the phase portrait depends sensitively on the behaviour of $G$ close to $r=0$. Only when the kernel $G$ is smooth at $r=0$ and the relative initial velocities of the particles are sufficiently large will the particles capture each other, otherwise they scatter. For the cases $k \leq 2$ the Green's function is not sufficiently smooth to typically allow capture orbits. 

Following this, in section~\ref{capture} we consider the case $k>2$, so that the kernel is sufficiently smooth to allow capture orbits for 2 particles, and show how the dynamics of a set of $N$ particles varies as the particles approach each other. We consider the behaviour of `dipoles', as we term pairs of particles that reach the limiting state of capture, and show examples of typical orbits for a group of $N=20$ particles over a number of timesteps. We conclude with a summary of the differences between the particle dynamics in $\diff(\R^2)$ with an $H^k$ metric, as described in this paper, and the point vortices for $\diff_{\rm vol}(\R^2)$ with an $L^2$ metric, and identify a number of questions that remain open.

\section{$N$ particle systems}

The Hamiltonian for the Euler equation (\ref{eq:Euler}) is the kinetic energy
\begin{equation}
H = \frac{1}{2} \int \langle\!\langle u,m \rangle\!\rangle \, dx dy = 
\frac{1}{2} \int m \mathcal{A}^{-1} m\, dx dy.
\end{equation}
Under the particle ansatz
\begin{equation}
m(x,t) = \sum_{i=1}^N p_i(t) \delta(x-q_i(t)), 
\label{eq:ansatz}
\end{equation}
($p_i(t), q_i(t) \in\R^{2}$ for each $i$)
the Hamiltonian becomes
\begin{equation}
H = \frac{1}{2}\sum_{i,j=1}^N p_i\cdot p_j G(\|q_i-q_j\|). \label{eq:Hamiltonian}
\end{equation}
Solutions to (\ref{eq:Euler}) of the form (\ref{eq:ansatz}) obey Hamilton's equations for 
(\ref{eq:Hamiltonian}), in which the components of $q_i$ and $p_i$ are canonically conjugate variables
(see~\cite{MarsdenRatiu99} for further details). Here $q_1,\dots,q_N$ represent the positions of the $N$ particles, and $p_1,\dots,p_N$ represent their momenta. 

There are two immediate differences from the analogous system for point vortices in the Euler fluid equations: the number of degrees of freedom is $2N$ instead of $N$, and, to get well-defined ODEs, the inertia operator $\mathcal{A}$ must be chosen so that $G(0)$ is finite. (For point vortices, $G(r)=-\log(r)/(2\pi)$, but the infinite self-energy of each particle can be ignored.)

The equations of motion are
\begin{eqnarray}
\label{eq:eom}
\dot q_i &=& \sum_{j=1}^N G(\|q_i-q_j\|)p_j, \nonumber \\
\dot p_i &=& -  \sum_{\scriptstyle j=1\atop \scriptstyle j\ne i}^N (p_i\cdot p_j)G'(\|q_i-q_j\|) \frac{q_i-q_j}{\|q_i-q_j\|}. \nonumber
\end{eqnarray}

They have
four conserved quantities:  the energy $H$, the linear momentum $\sum_{i=1}^N p_i$, and the angular momentum $\sum_{i=1}^n q_i\times p_i$.  These are sufficient to ensure the integrability of the two particle problem, which is discussed in section~\ref{sec:2body}. However, the dynamics of these two-body problems are very different from more familiar ones on flat configuration spaces, like the Kepler problem. Also note that Eqs. (\ref{eq:eom}) are not Galilean-invariant; their dynamics depends on the linear momentum.

\section{Blobs and kernels \label{findkernel}}
The Euler equation for $\diff(\R^n)$ and inertia operator $\mathcal{A}$ is given by (\ref{Euler1}) and (\ref{Euler2}). A related equation, the {\it regularized Euler equation}, is given by (\ref{Euler1}) together with $u=\mathbf{G}_\psi*m$ (in place of (\ref{Euler2})), where $\mathbf{G}_\psi$ is the regularized Green's function
$$\mathbf{G}_\psi(x,x') = \int G(x,x')\psi(x')\, dx'.$$
Typically, $\psi(x) = f(\|x\|)$ is a `blob', or approximate delta-function.

For a given inertia operator ${\mathcal A}$, the kernel $G$ is  the velocity corresponding to a blob $f(\|x\|)$ of momentum. It satisfies
$${\mathcal A}G = f(\|x\|),$$
which (assuming $\mathcal A$ is rotationally invariant) can be solved using the Hankel transform as
$$u(x) = \int_0^\infty \frac{\tilde f(m)}{\tilde a(m)}J_0(m\|x\|)m\, dm, $$
where
$$\tilde f(m) = \int_0^\infty f(r) J_0(m r) r \, dr$$
is the Hankel transform of $f$, and $\tilde a(m)$ is the Fourier symbol of the operator $\mathcal A$, e.g., $\tilde a(m) = 1+m^2$ for $\mathcal A = 1-\nabla^2$.

Notice that considering a regularised Euler inertia operator with symbol $\tilde a(m)$ together with blobs of symbol $\tilde f(m)$ is identical to considering the standard Euler inertia operator with symbol $\tilde a(m)/\tilde f(m)$ together with delta-function particles. 

Consider first the $H^2$ metric associated with the inertia operator $\mathcal{A}_2 = (1-\nabla^2)^2$. Its kernel is $G(r) = - r K_1(r)/(4\pi)$, where $K_1(r)$ is a modified Bessel function of the first kind. Note that $G(r)$ is negative for all $r$, reaching its global minimum ($-1/(4\pi)$) at $r=0$. Thus, the velocity field induced by a single such particle is in the direction $-p$. We have found it more convenient to change the sign of all kernels, so that the induced velocity field is in the direction $+p$. Since the dynamics are reversible, this does not materially affect them. In this case, we work with $G(r) = r K_1(r)/(4\pi)$.

\begin{figure}
\begin{center}
\includegraphics[width=10cm]{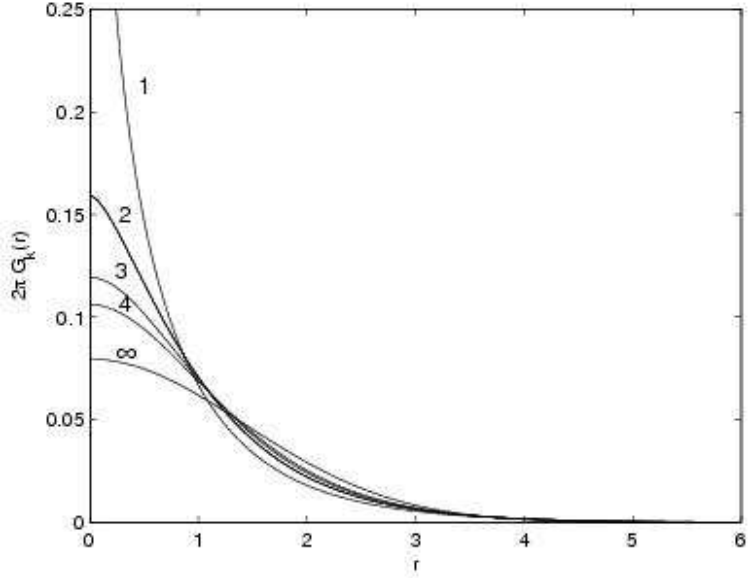}
\caption{Kernels $G_k(r)$ found in Section \ref{findkernel} for a family of $H^k$ metrics ($k = 1, 2, 3, 4, \infty$) with delta-function momentum. }
\label{kernels}
\end{center}
\end{figure}

For numerical simulations it is convenient to have an explicit form for $G(r)$. In addition, it turns out that the dynamics of the 2-body problem depends sensitively on $G(r)$ near $r=0$. Therefore, we have worked with the family of $H^k$ metrics with inertia operators ${\mathcal A}_k  := (1- \alpha^2 \nabla^2)^k$ and delta-function particles. Then the kernels $G_k$ can be found explicitly in terms of modified Bessel functions. The first few are (in terms of the scaled length $\tilde r := r/\alpha$)
 \begin{eqnarray}
 2\pi \alpha^{2} G_1(r) &=&K_0(\tilde r) \nonumber \\ 
 &=& \log(2/\tilde r)- \gamma
 + \O(r^2\log r),  \nonumber \\
 2\pi \alpha^{2} G_2(r) &=& \frac{1}{2} \tilde r K_1(\tilde r)\nonumber \\
 &=& \frac{1}{2}-\frac{1}{8}\left( 2\log(2/\tilde r) -2\gamma+1 \right) {\tilde r}^{2}-{\frac {1}{128}}\left( 4\log(2/\tilde r)+5-4\gamma \right) {\tilde r}^{4}+O \left( {r}^{6} \right),\nonumber  \\
\hbox{\rm and\ }
  2\pi \alpha^{2} G_3(r) &=& 
  \frac{1}{8}\tilde r \left( \tilde rK_0(\tilde r) +2 K_1(\tilde r) \right)\nonumber  \\
&=& \frac{1}{4}-\frac{1}{16}{\tilde r}^{2}+{\frac {1}{256}}\left( 4\log(2/\tilde r)+3-4\gamma \right) {\tilde r}^{4}+O \left( {r}^{6} \right). \nonumber 
\end{eqnarray}
Choosing the length scale $\alpha = \eps/ \sqrt{k}$ we have formally, 
$${\mathcal A}_k\to{\mathcal A}_\infty := \exp(-\eps^2\nabla^2),$$
with Fourier symbol $\tilde a(m) = \exp(-\eps^2 m^2)$ and Green's function
$$2 \pi G_\infty(r) = \frac{1}{2\eps^2}\exp\left(-\frac{r^2}{4\eps^2}\right).$$
That is, the Gaussian kernel is Green's function for an $H^\infty$ metric, the limit of a family of $H^k$ metrics. The family of kernels is shown in Figure~\ref{kernels}.

\section{2 particle systems \label{sec:2body}}

In this section we derive the reduced Hamiltonian for the case where there are only 2 particles. This enables us to describe the phase portraits, and to consider how they depend on the behaviour of the kernel, and hence on the metric. 

\subsection{The reduced Hamiltonian \label{sec:reduction}}

Let $N=2$, so that the Hamiltonian (see Eq.~\ref{eq:Hamiltonian}) is
$$H =  p_1\cdot p_2 G(\|q_1-q_2\|) + \frac{1}{2} (\|p_1\|^2 + \|p_2\|^2)G(0).$$
We first change to the canonical Jacobi-Haretu centre of mass coordinates~\cite{MarsdenRatiu99} in order to reduce by the translational symmetry. Let 
$$d = p_1 + p_2,\quad P = (p_1-p_2)/2,\quad  c = (q_1+q_2)/2,\quad Q = q_1-q_2,$$
or
$$p_1 = \frac{1}{2}d +P,\quad p_2 = \frac{1}{2}d - P,\quad q_1 = c + \frac{1}{2}Q, \quad
q_2= c - \frac{1}{2}Q.$$
Then
$$H = \left(\frac{1}{4}\|d\|^2-\|P\|^2\right)G(Q) + \left(\frac{1}{4}\|d\|^2 + \|P\|^2\right)G(0).$$
Changing to symplectic polar coordinates with 
$$Q=(r\cos\theta,r\sin\theta),\quad
P=(p\cos\theta-p_\theta\sin\theta/r,p\sin\theta+p_\theta\cos\theta/r),$$
where $(r,p)$ and $(\theta,p_\theta)$ are conjugate variables,
the reduced Hamiltonian becomes 
\begin{equation}H=\frac{1}{4}\|d\|^2(G(0)+G(r)) +   \left( p^2+\frac{p_\theta^2}{r^2} \right) (G(0)-G(r)).\label{Hred}
\end{equation}
The conserved quantities are the momentum of the centre of mass, $d$, and the angular momentum $p_\theta =: \mu$. 
The phase portrait of this reduced Hamiltonian depends very sensitively on the behaviour of $G$ near $r=0$.

\begin{figure}
\begin{center}
\includegraphics[width=8cm]{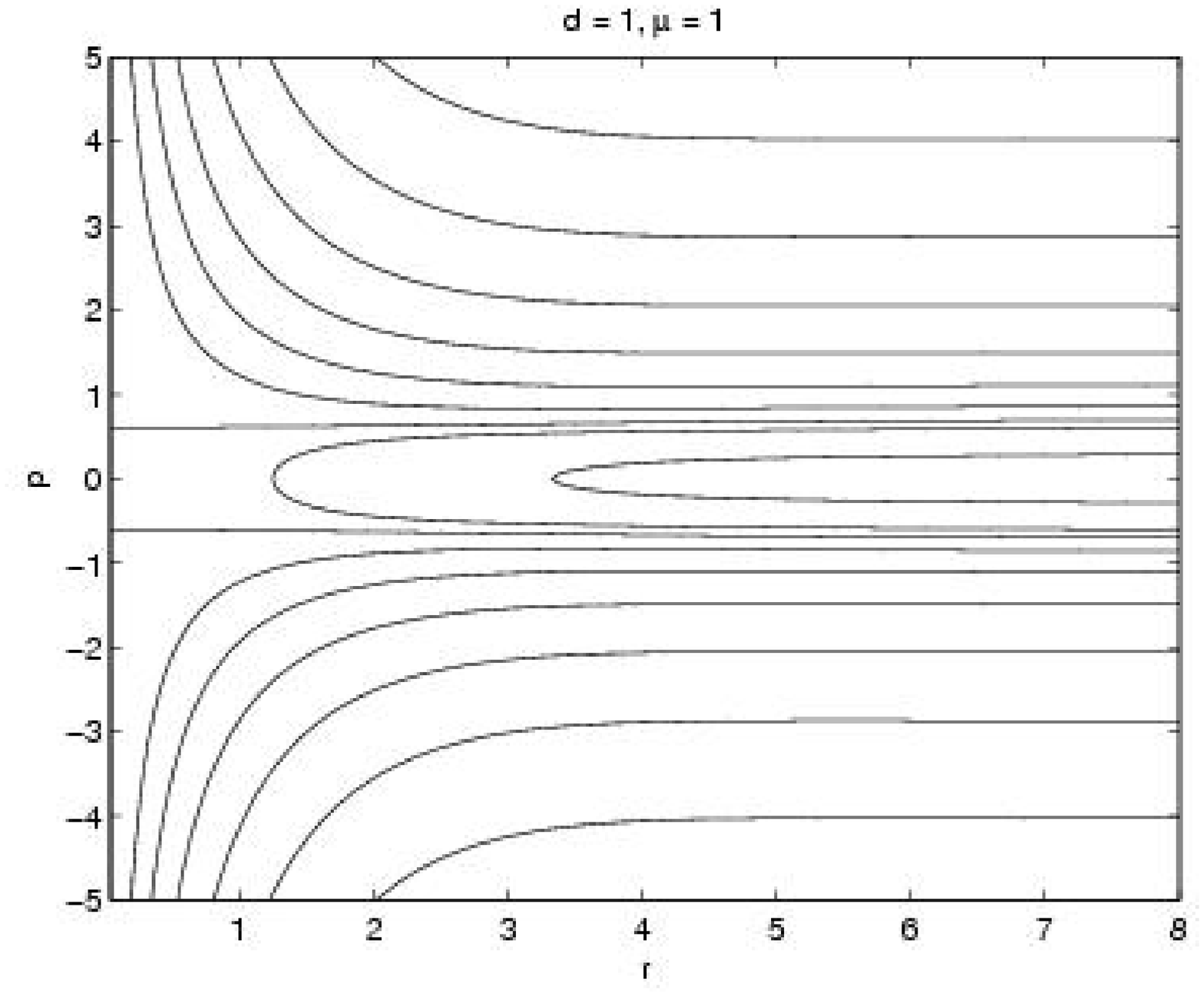}
\includegraphics[width=8cm]{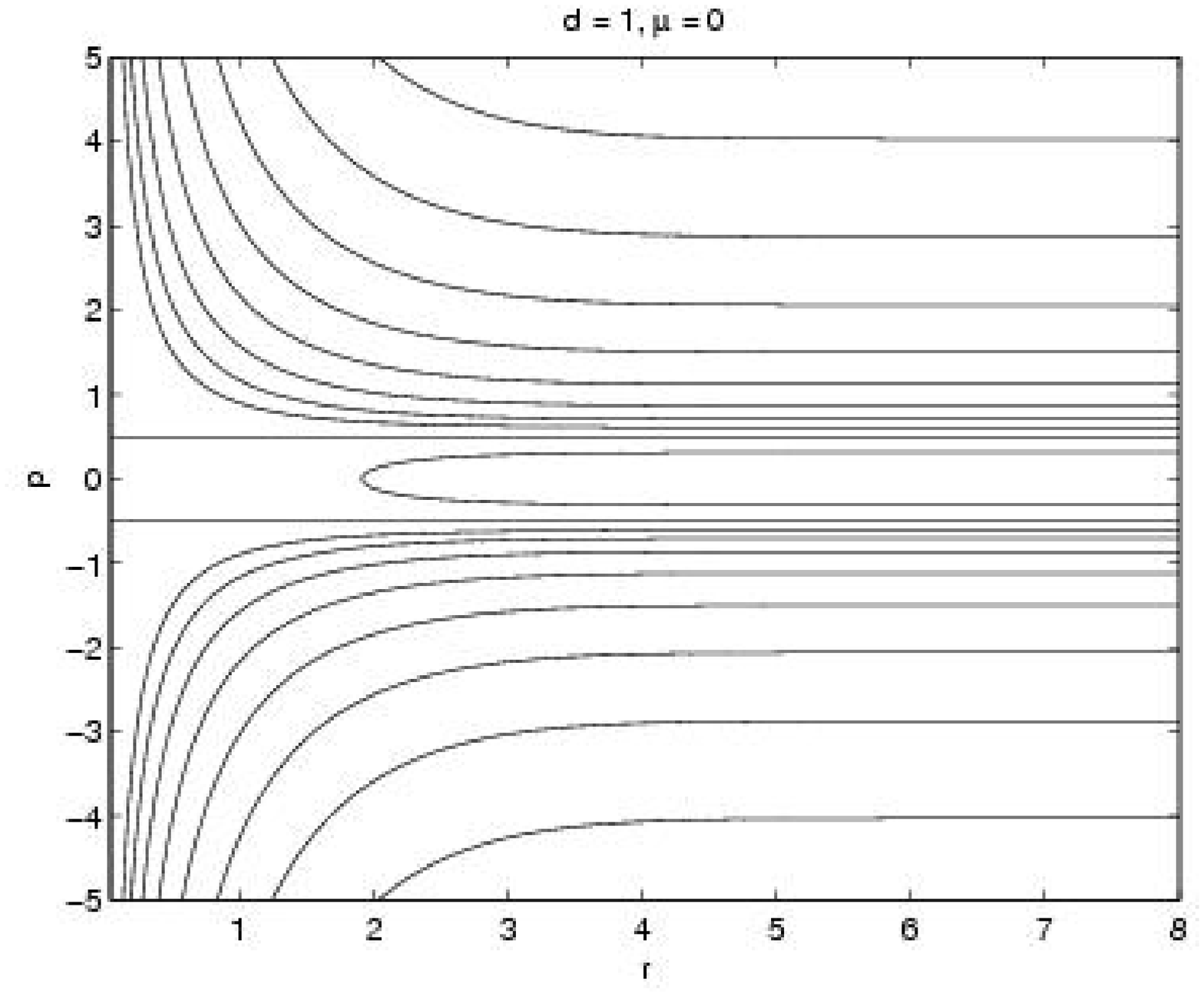}
\includegraphics[width=8cm]{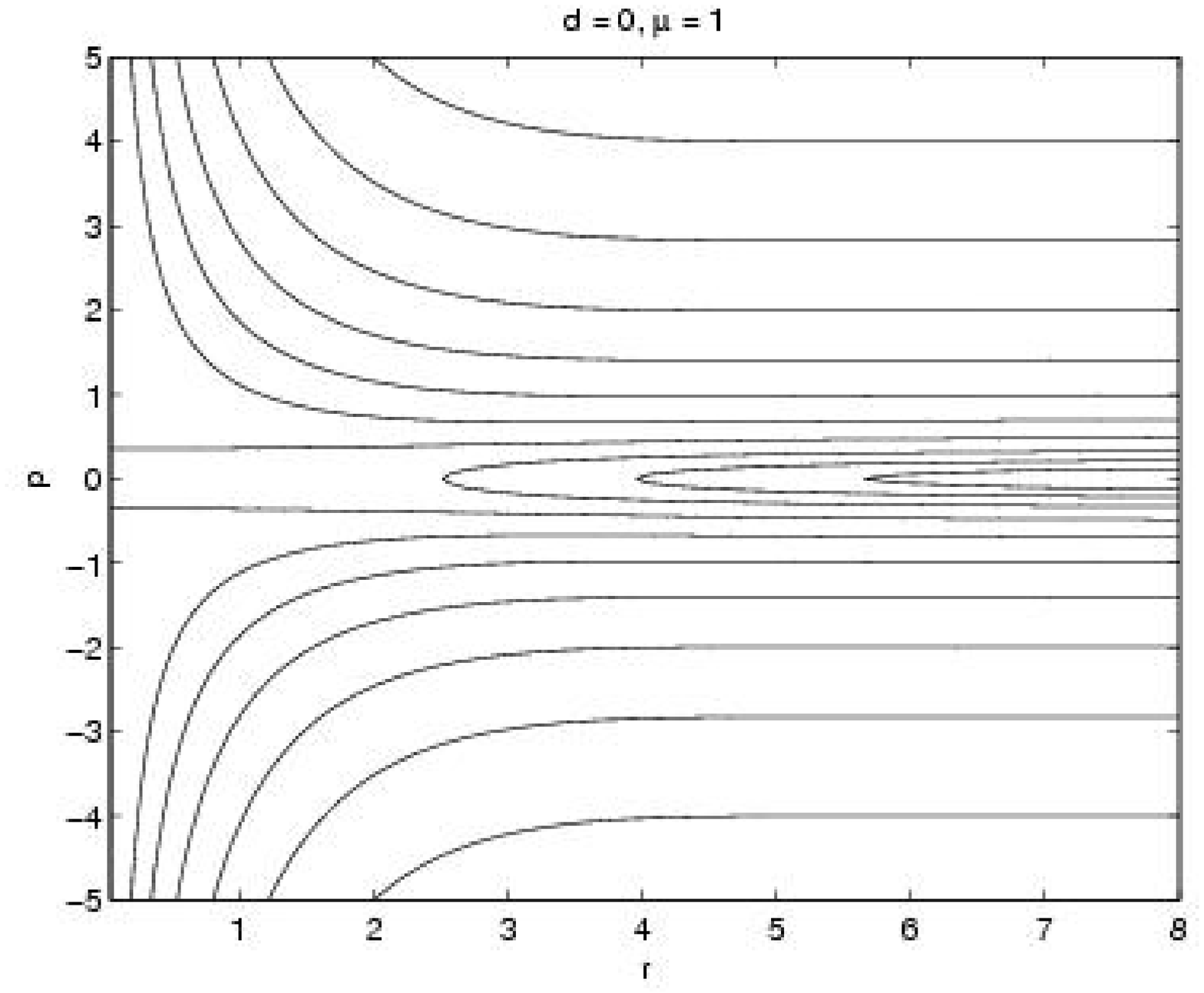}
\includegraphics[width=8cm]{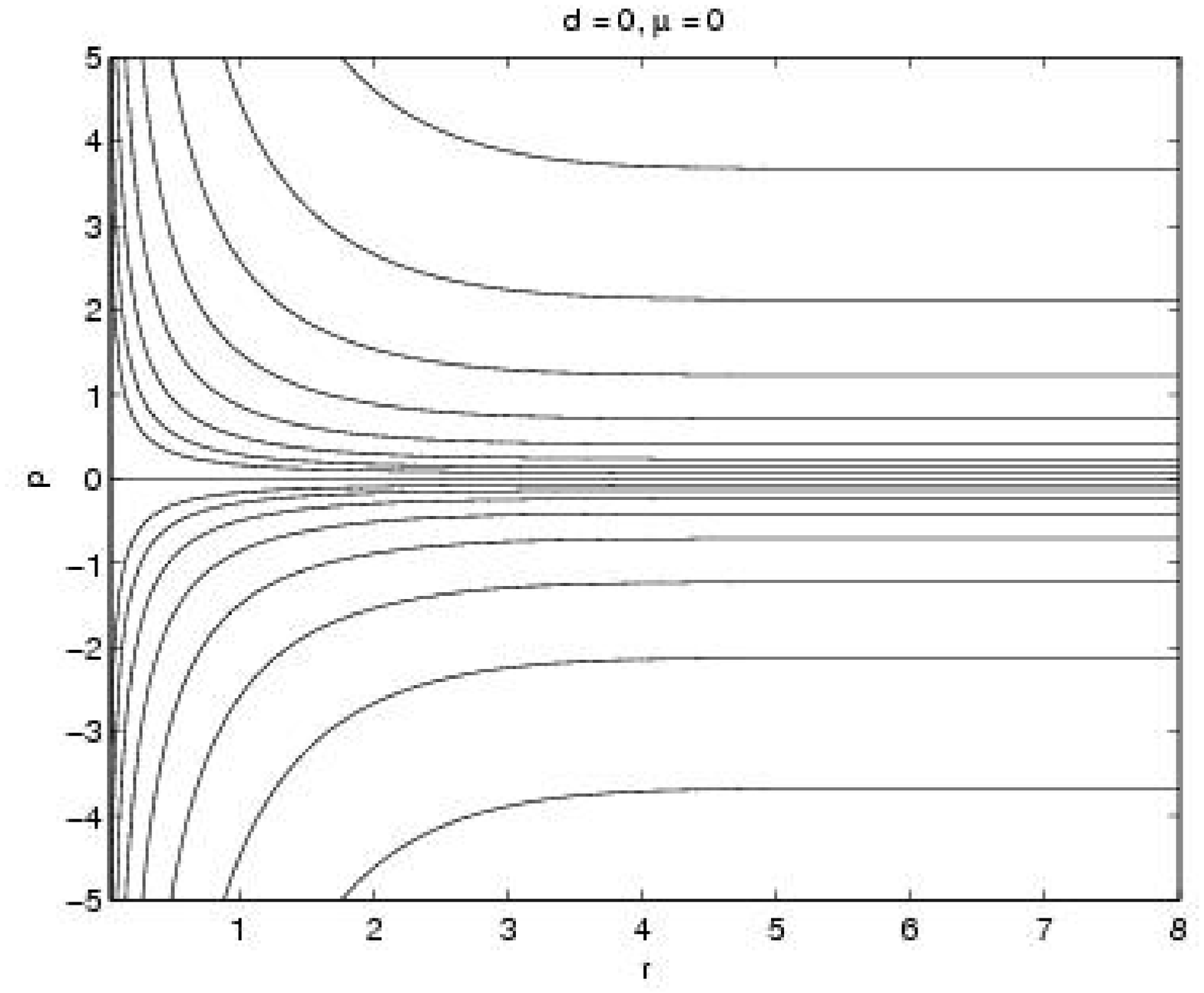}
\caption{Phase portrait for the 2 body problem with Gaussian kernel $G(r)=\exp(-r^2/4)$. The Hamiltonian is given by Eq. (\ref{Hred}). Top: Nonzero linear momentum $d$. Bottom: zero linear momentum. Left: Nonzero angular momentum $\mu$. Right: Zero angular momentum, for which the motion is confined to a line. \label{phase3}}
\end{center}
\end{figure}

First, consider the case when $G$ is smooth at $r=0$, so that  $G(r)= G(0) + \frac{1}{2}G''(0) r^2
+ \frac{1}{24}G^{(4)}(0)r^4+\O(r^6).$ In this case, near $r=0$, the Hamiltonian is
\begin{equation}
\label{Happrox}
H = \frac{1}{2}\left(G(0)\|d\|^2 - G''(0)\mu^2\right)+
\frac{1}{24} \left(3G''(0)(\|d\|^2 - 4 p^2)-G^{(4)}(0)\mu^2\right)r^2  + \O(r^4).
\end{equation}
Therefore, a separatrix intersects the $p$-axis at 
$$p^2=\frac{1}{4}\|d\|^2-\frac{G^{(4)}(0)}{12 G''(0)}\mu^2.$$
(Typically, and in all the examples of this paper, $G''(0)<0$ and $G^{(4)}(0)>0$.)
The energy on the $p$-axis, and hence on the separatrix, is $\frac{1}{2}\left(G(0)\|d\|^2 - G''(0)\mu^2\right)$. As $r\to\infty$, $G(r)\to 0$ so $H\sim G(0)(p^2 + \frac{1}{4}\|d\|^2)$ and the separatrix has a horizontal asymptote at
\begin{equation}
p^2 = \frac{1}{4}\|d\|^2 - \frac{G''(0)}{2G(0)}\mu^2. \label{separatrix}
\end{equation}
This can be clearly seen in Figure~\ref{phase3}, where the phase portrait is given for the 2 body problem under the ${\mathcal A}_{\infty}$ Gaussian kernel. The phase portraits of the 2 body problem under other metrics are given in Figures~\ref{phase2} and~\ref{phase1} and discussed in section~\ref{sec:dynamics}.

Thus there are three principal behaviours of pairs of particles:
\begin{description}
\item[Scattering] Both particles are free as $t \to \pm \infty$
\item[Capture] Distance between particles $r \to 0$ as $t \to \infty$
\item[Ejection] Distance between particles $r \to 0$ as $t \to -\infty$
\end{description}
as well as degenerate capture/ejection orbits given by the separatrices.
Inside the separatrix  the two particles scatter, while outside the separatrix they either capture ($pG''(0)<0$) or eject ($pG''(0)>0$) each other.

We now consider the criterion for a capture to take place.
Equation (\ref{separatrix}) tells us that two particles initially very far apart will capture each other if $p^2 > \frac{1}{4}\|d\|^2 - \frac{G''(0)}{2G(0)}\mu^2$. This criterion, depending on the three parameters $p$, $\|d\|$, and $\mu$, can be expressed more intuitively as follows. We introduce the angle of approach $\psi := \angle p_1 p_2$, the speed ratio $\|p_1\|/\|p_2\|$, and the separation distance $D$. $D$ is the distance of closest approach of two particles initially at $q_{1,2}$ moving in straight lines with direction $p_{1,2}$. Clearly, as $D$ increases then the particles cease to interact, so we will seek the critical value of $D$ for a capture to occur.

\begin{figure}
\begin{center}
\includegraphics[width=8cm]{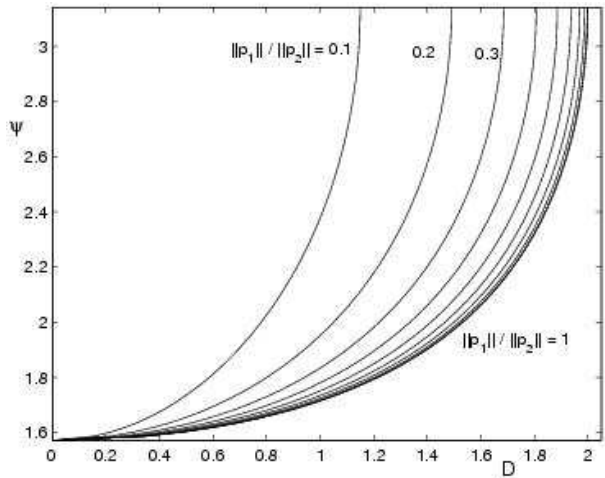}
\caption{Minimum required approach angle $\psi\in(\frac{\pi}{2},\pi)$ for a capture to take place for a pair of initially distant particles with speeds $\|p_{1,2}\|$ and separation distance $D$. Here we have taken $G(r)=e^{-r^2/4}$, for which the maximum possible separation distance in a capture
is 2. That this equals the range of $G$ is something of a coincidence, because 
the critical $D$ depends only on $G(0)$ and $G''(0)$.
\label{fig:captures}
}
\end{center}
\end{figure}

We have $D = \|Q\|\sin(\angle QP)$ and $\mu = Q\times P = \|Q\| \|P\| \sin(\angle QP) =
D\|P\|$. As $r\to\infty$, $p^2 \to \|P\|^2 = \frac{1}{4}\|p_1-p_2\|^2$, so the capture criterion becomes
$$ -p_1\cdot p_2 = - \|p_1\| \|p_2\| \cos\psi > -\frac{G''(0)}{2G(0)} D^2 \|P\|^2,$$
or, using the law of cosines $\|2P\|^2 = \|p_1-p_2\|^2 = \|p_1\|^2 + \|p_2\|^2 - 2 \|p_1\| \|p_2 \| \cos\psi$,
$$ \cos\psi < \frac{\widetilde D}{1+2\widetilde D}\left(\frac{\|p_1\|}{\|p_2\|} + \frac{\|p_2\|}{\|p_1\|}\right),
\quad -\frac{1}{4} < \widetilde D < 0,$$
where
$$\widetilde D = \frac{G''(0)}{8 G(0)} D^2.$$
That is, a separation distance $D$ of less than $\sqrt{-2G(0)/G''(0)}$ is necessary for a capture
to be possible. Particles with $D=0$ (equivalently, $\mu=0$) are captured when $|\psi|>\pi/2$
(a `head-on' approach) and scattered when $|\psi|<\pi/2$ (a `glancing' approach), for
all $\|p_1\|/\|p_2\|$; increasing $D$ and $\|p_1|/\|p_2\|$ moving away from 1 both restrict
captures to more nearly head-on approaches. Figure \ref{fig:captures} shows the minimum required approach angle $\psi$ for each $D$.

During a capture the particles get infinitely close, $r\to 0$. Although $p\to-\infty$, the actual speed of approach of the particles, $\dot r$, tends to 0. The particles do not collide in a finite time but rather become bound together, or stick to each other. For large $p$ and small $r$, we can use the approximate form of the Hamiltonian for $r \approx 0$ to get the approximate equations of motion:
$$\dot r = -G''(0)p r^2,\quad \dot p =  G''(0) p^2 r,$$
with solution
\begin{equation}r(t) = r_0 e^{-\alpha t},\quad p(t) = p_0 e^{\alpha t},\label{eq:motion}\end{equation}
where the constant $\alpha =  G''(0) p_0 r_0$, corresponding to a capturing orbit for $p_0G''(0)<0$ and an ejecting orbit for $p_0G''(0)>0$. We call a pair of particles in the limiting state of (\ref{eq:motion}) a dipole, and explore them further in section~\ref{capture}.

For a scattering orbit, using $$\dot\theta = \frac{\partial H}{\partial p_\theta} = 2\mu \frac{G(0)-G(r)}{r^2},$$
we see that the scattering angle $\theta|_{-\infty}^{\infty}$ ranges over $(0,\infty)$ as the minimum separation of the particles decreases from $\infty$ to 0. (As $r\to0$, $\dot\theta=\O(1)$ and  $\dot p = \O(r)$, so  the phase $\theta$ accumulates without bound.) Again, because close particles move very slowly, there can be an arbitrarily long delay during the close approach---the particles appear to stick together for a while.

\subsection{Particle dynamics under different metrics \label{sec:dynamics}}

We can consider different kernels (i.e., different functions $G(r)$) and use them to discuss the particle dynamics for each metric in turn. In all cases, we consider the Helmholtz-style inertia operator described previously: ${\mathcal A}_k  := (1- \alpha^2 \nabla^2)^k$. We begin with the $H^1$ metric, and progress through the different $H^k$ metrics.

Delta-function particles for the inertia operator ${\mathcal A}_1$  ($H^1$ metric) are not well-defined because they do not have a finite self-induced velocity (the same situation occurs in the 3D Euler fluid equations when the vorticity is concentrated on a curved filament). To use the delta-function particles, it is necessary to regularize the equations. One approach to this would be to simply set $G(0) = 0$, so that the self-induced velocity of each particle is 0. We consider this below (section~\ref{g00}), but we first consider the standard regularization of introducing a smooth blob function $f(x)$, so that we consider the operator $\mathcal{A} = \mathcal{A}_1/f(x)$, as we described earlier.

As mentioned in the introduction, we consider the Gaussian blob $f(x) = \frac{1}{\pi}e^{-x^2/\eps^2}/\eps^2$, which tends to $\delta(x)$ as
$\eps\to 0$. In this case, $\tilde f(m) = \frac{1}{2\pi}e^{-\eps^2 m^2/4}$. The kernel for this blob is smooth with the inertia operator $\mathcal A_1 = 1-\nabla^2$, so we apply the results of section~\ref{sec:reduction} and expand for small $x$ as:
$$2\pi G(r) = C + \left(-\frac{1}{2\eps^2} + \frac{1}{4}C\right)x^2
+ \left(\frac{1}{8\eps^4} - \frac{1}{32\eps^2}+\frac{1}{64}C
\right) x^4 + \O(x^6),$$
where $C = \frac{1}{2}\exp(\frac{1}{4}\eps^2)\mathop{\rm Ei}_1(\frac{1}{4}\eps^2)$.

As $\eps\to 0$ we have $G(0)\sim (-2\log\eps - \gamma + \log 4)/(4\pi)$,
$G''(0)\sim -1/(2\pi\eps^2)$, and $G''''(0)\sim 3/(2\pi\eps^4)$, and so the separatrix intersects
$r=0$ at $p = \O(\mu/\eps)$, with a horizontal asymptote
 at $p = \O(\mu/(\eps(\log|\eps|)^{1/2})$. As $\eps\to 0$, providing that $\mu \ne 0$, the separatrix moves outward and particle captures become rarer. 
 
\begin{figure}
\begin{center}
\includegraphics[width=8cm]{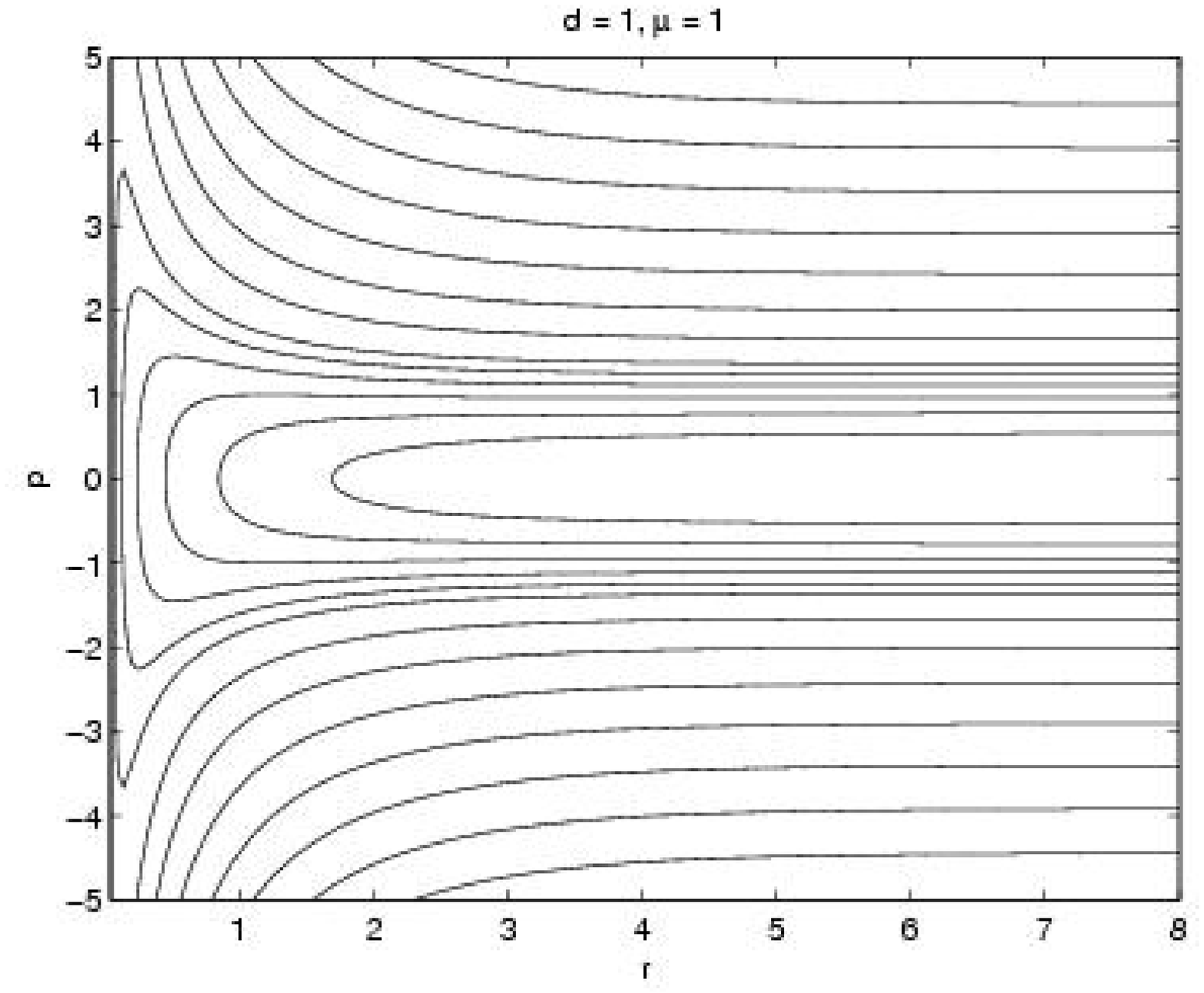}
\includegraphics[width=8cm]{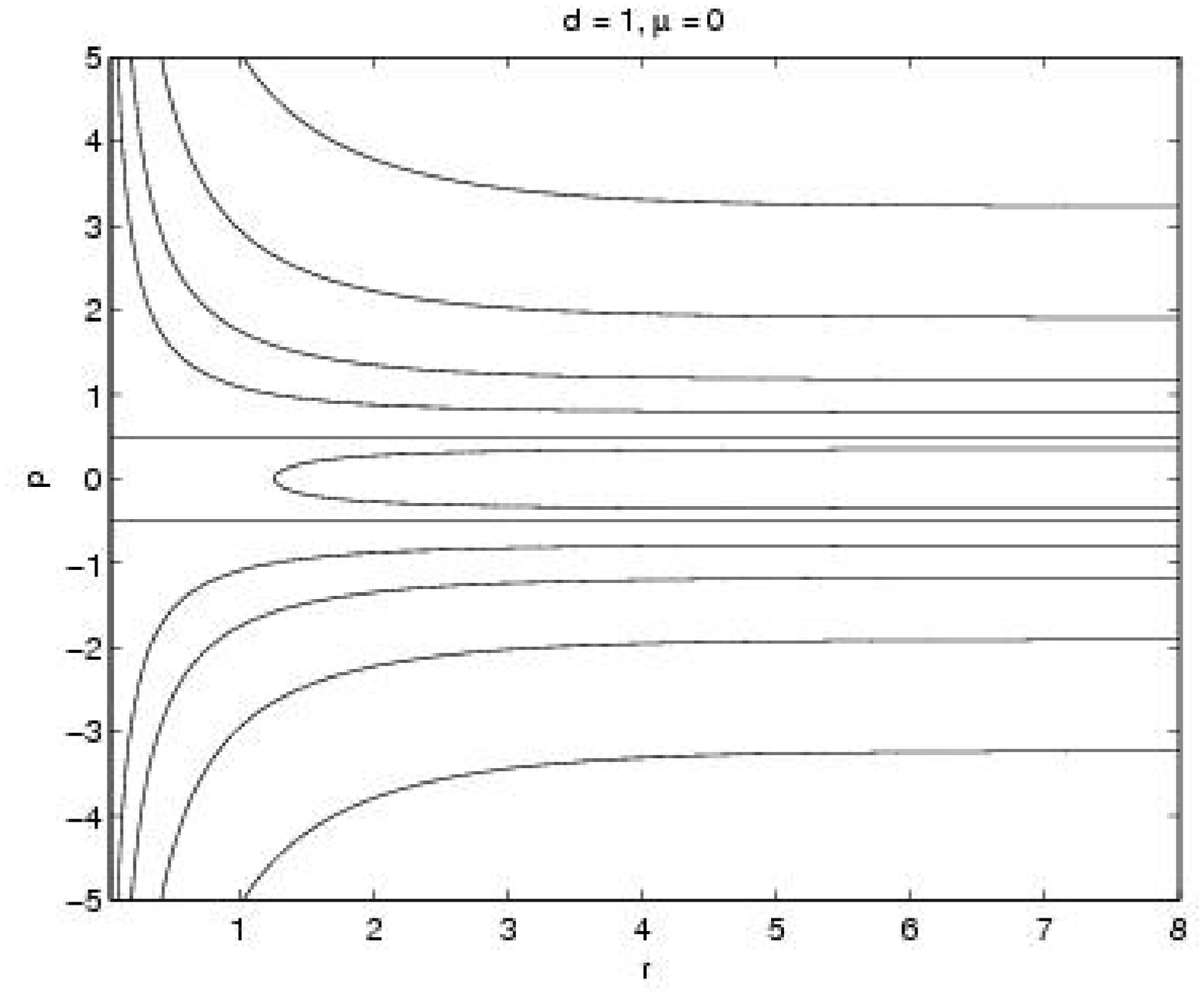}
\includegraphics[width=8cm]{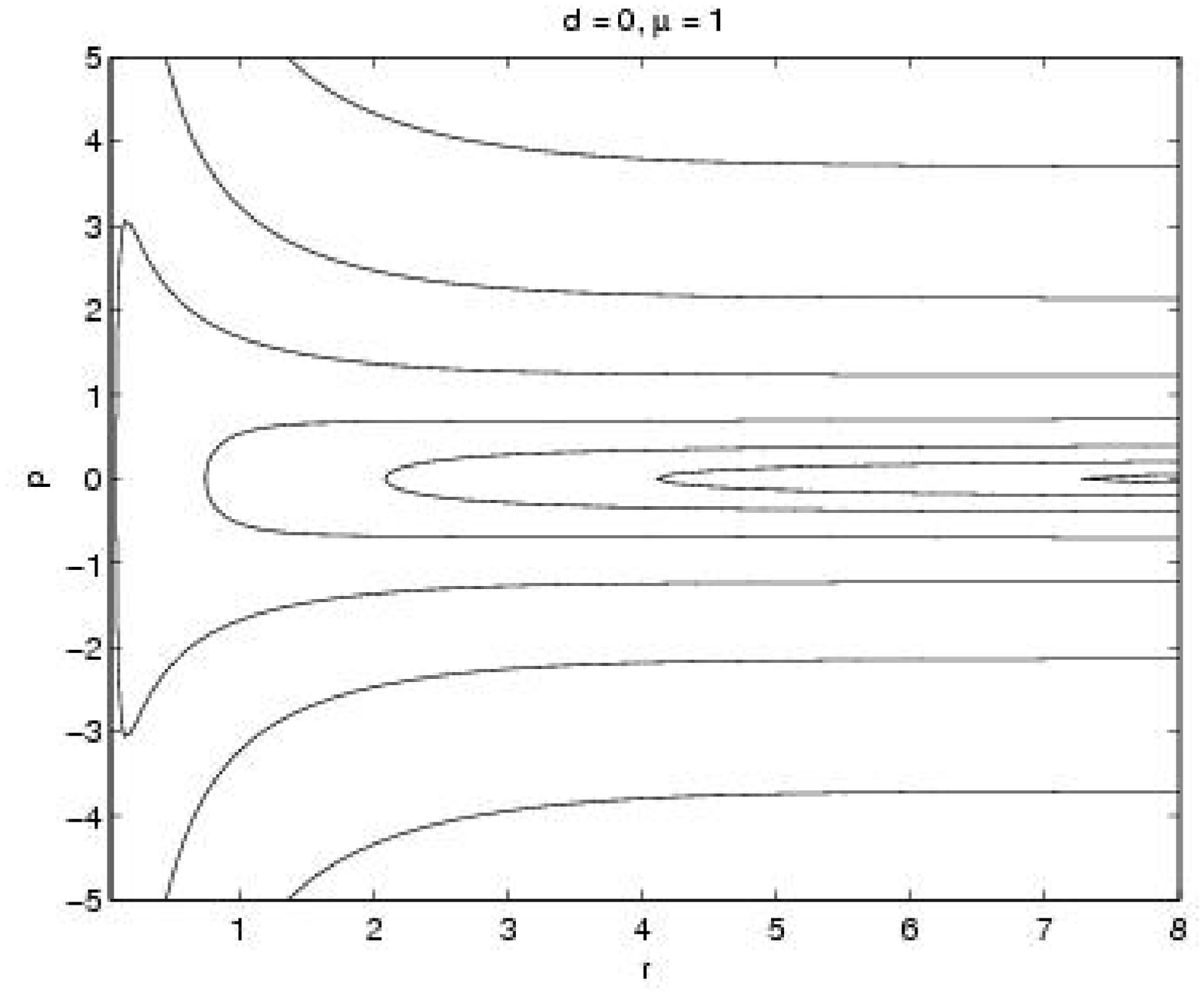}
\includegraphics[width=8cm]{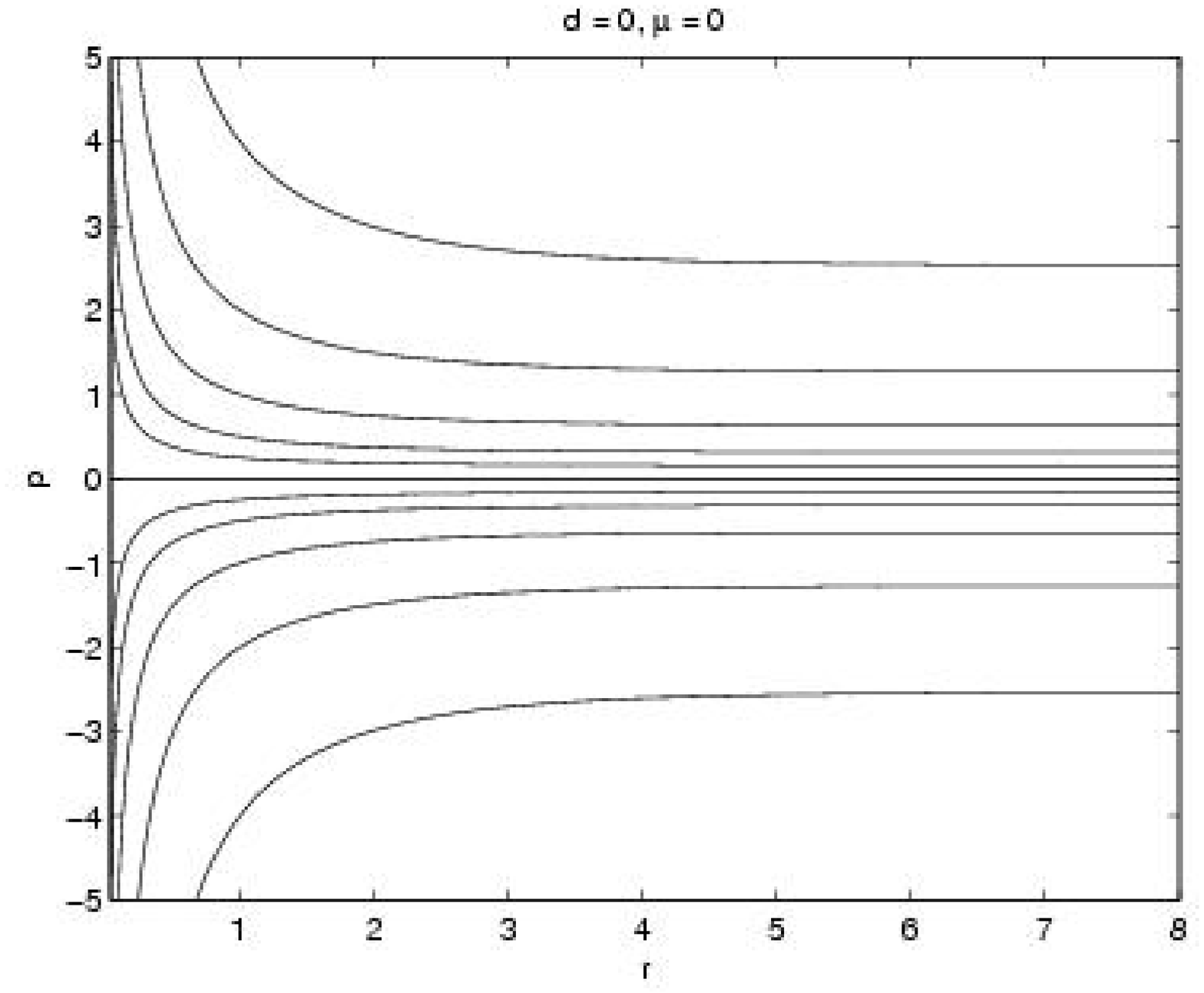}
\caption{Phase portrait for the 2 body problem with inertia operator ${\mathcal A}_2=(1-\nabla^2)^2$. The Hamiltonian is given in Eq. (\ref{Hred}). Unless the angular momentum $\mu$ is zero (in which case the motion is confined to a line), all orbits are scattering.}
\label{phase2}
\end{center}
\end{figure}

\begin{figure}
\begin{center}
\includegraphics[width=12cm]{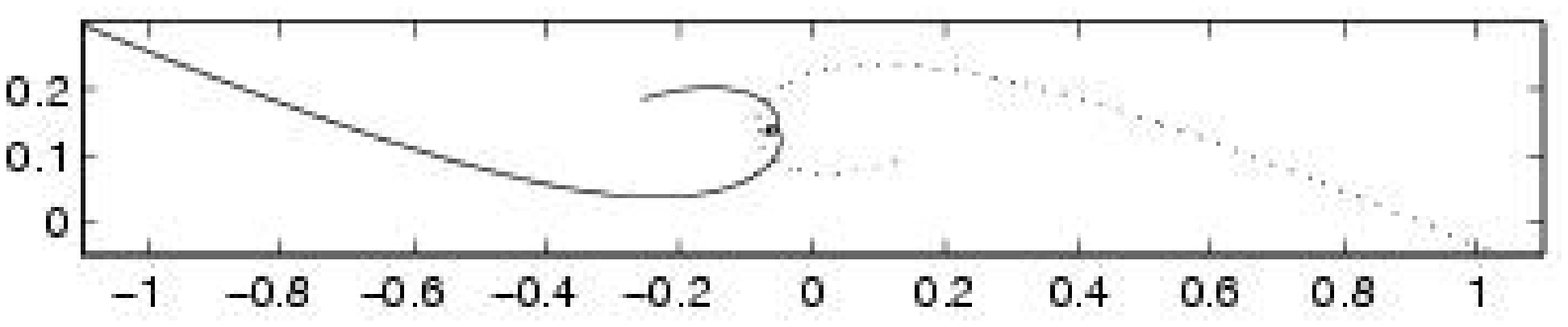}
\includegraphics[width=8cm]{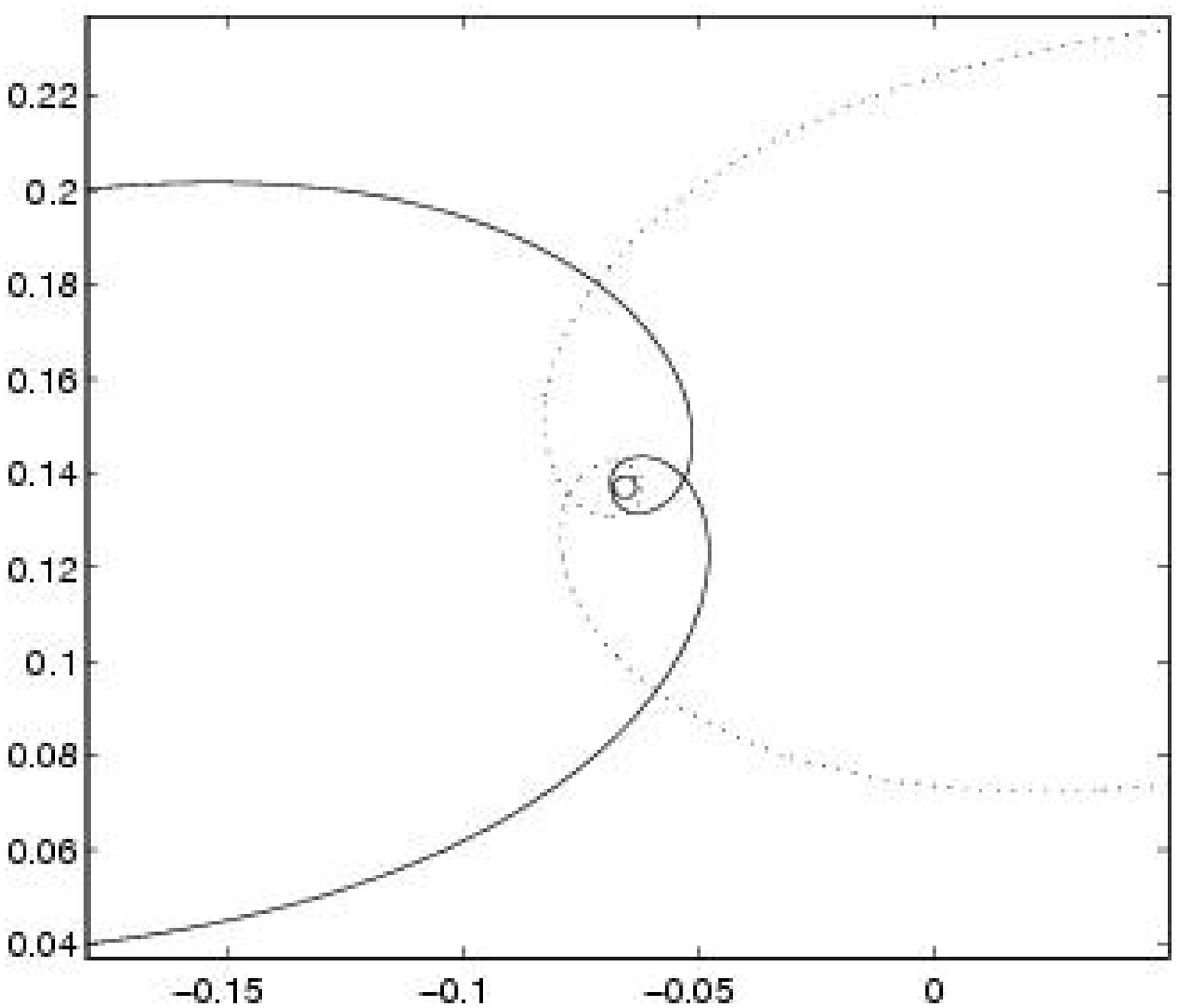}
\caption{A typical scattering orbit for the 2 body problem with inertia operator ${\mathcal A}_2$. The relative momentum $2P=p_1-p_2$ can become arbitrarily large during close approaches.}
\label{dance2}
\end{center}
\end{figure}

The $H^2$ metric is smooth enough for delta-function particles to be well-defined, with a finite self-induced velocity. The  phase portrait for this case is shown in Figure. \ref{phase2}. However, $G_2(r)$ is not twice differentiable at $r=0$ and so the above analysis of the phase portrait of the two body problem is not valid. In fact, the singularity of $G_2(r)$ is sufficient to prevent the occurrence of capture orbits; when $\mu\ne 0$ all orbits are scattering (see Figure~\ref{phase2}). However, two particles with large initial relative momentum $p$ can approach arbitrarily closely, and achieve arbitrarily high $p$ during the close encounter (which makes them behave a lot like capturing orbits); but they do eventually separate. A typical such encounter is shown in Figure~\ref{dance2}.

The $H^3$ metric is smoother, but $G_3(r)$ is still not 4 times differentiable at $r=0$, so the above separatrix analysis is still not valid. However, one can check that the 2 body problem does now have a separatrix that divides capture and scattering orbits, although it is not a smooth curve at $r=0$. 

\begin{figure}
\begin{center}
\includegraphics[width=8cm]{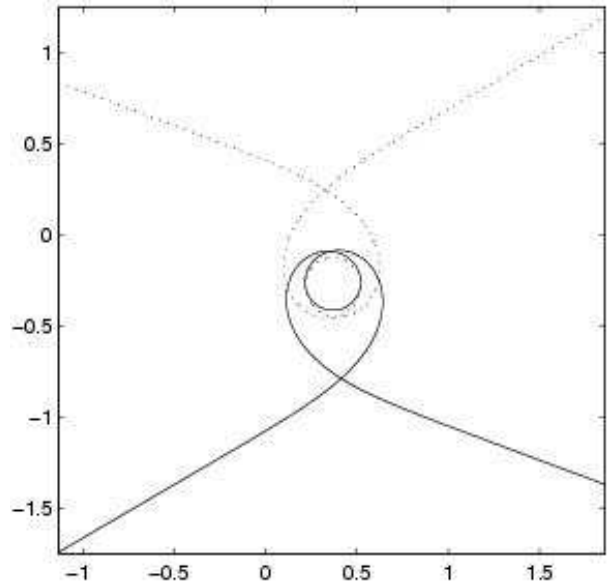}
\caption{A typical scattering orbit for the 2 body problem with inertia operator ${\mathcal A}_\infty$. The relative momentum $2P=p_1-p_2$ remains bounded during the close approach.\label{dance3}}
\end{center}
\end{figure}

For $H^k$ ($k\ge 4$) and $H^\infty$, the separatrix is as described above. Figure~\ref{phase3} shows the phase portrait for the $H^{\infty}$ metric. A typical scattering orbit is shown in Figure~\ref{dance3}; in contrast to the $H^2$ case, the relative momentum $2P=p_1-p_2$ is bounded over all scattering orbits.

In no case are there any periodic orbits in the two body problem.

\subsection{An alternative regularization for $H^1$ \label{g00}}

As was mentioned above, there is another possible regularization of the $N$-particle problem with the $H^1$ metric, which is simply to set the self-induced velocity of each particle to zero, i.e., $G(0)=0$. This seems drastic, when the `correct' self-induced velocity of a delta-function is infinite, but it still corresponds to a consistent discretization of the PDE (\ref{eq:Euler}) in the limit of a large number of particles spaced over a curve or an area. (The contribution to the velocity at a point from nearby momentum $p$ is $\int_0^\eps p(r) G_1(r) r\, dr$ for a curve, and $\int_0^\eps\!\!\int_0^{2\pi} p(r,\theta) G_1(r) r^2\, dr$ for an area; in both cases the logarithmic singularity in $G_1(r)$ is weak enough that the nearby momentum does not contribute.) 

\begin{figure}
\begin{center}
\includegraphics[width=8cm]{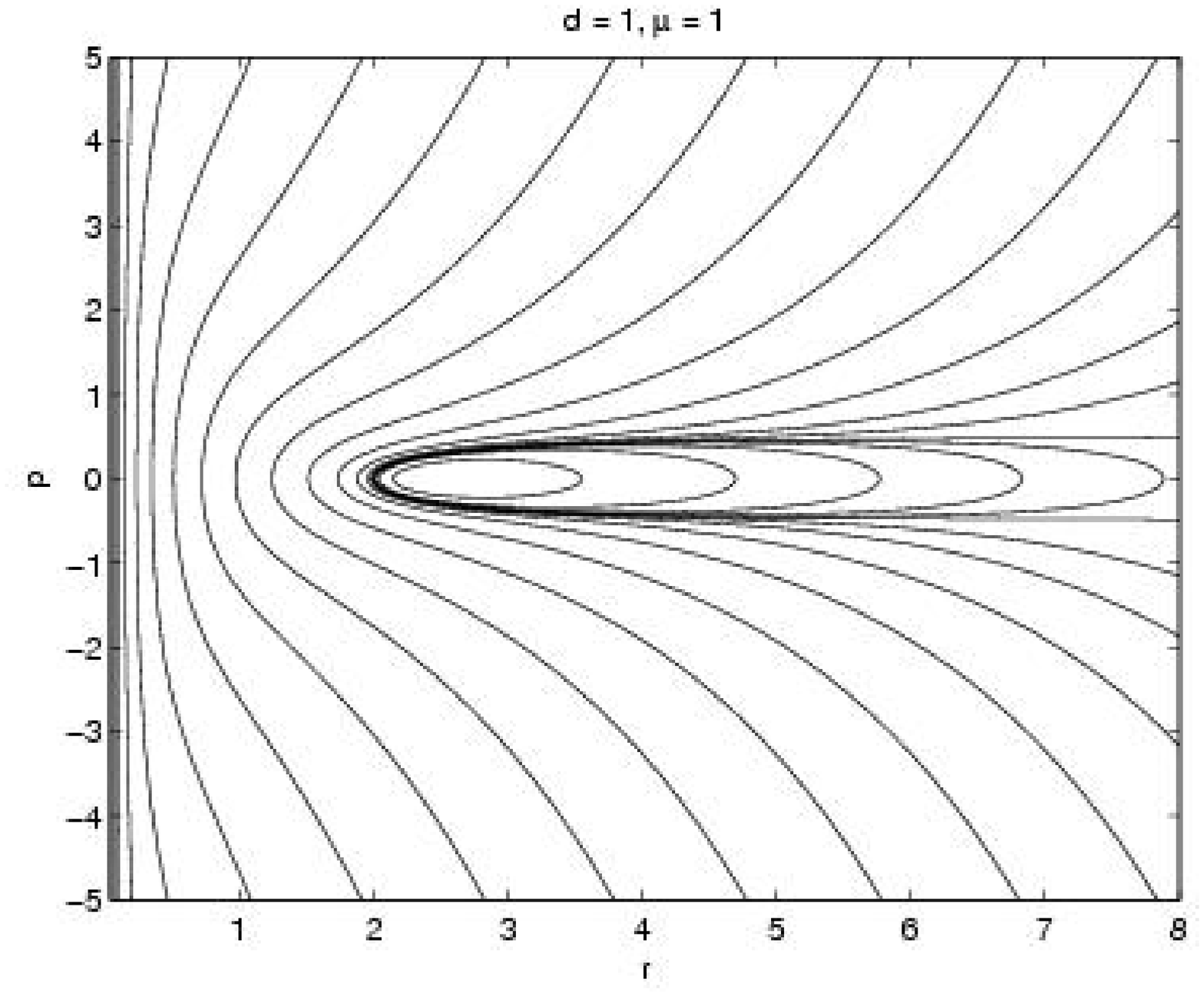}
\includegraphics[width=8cm]{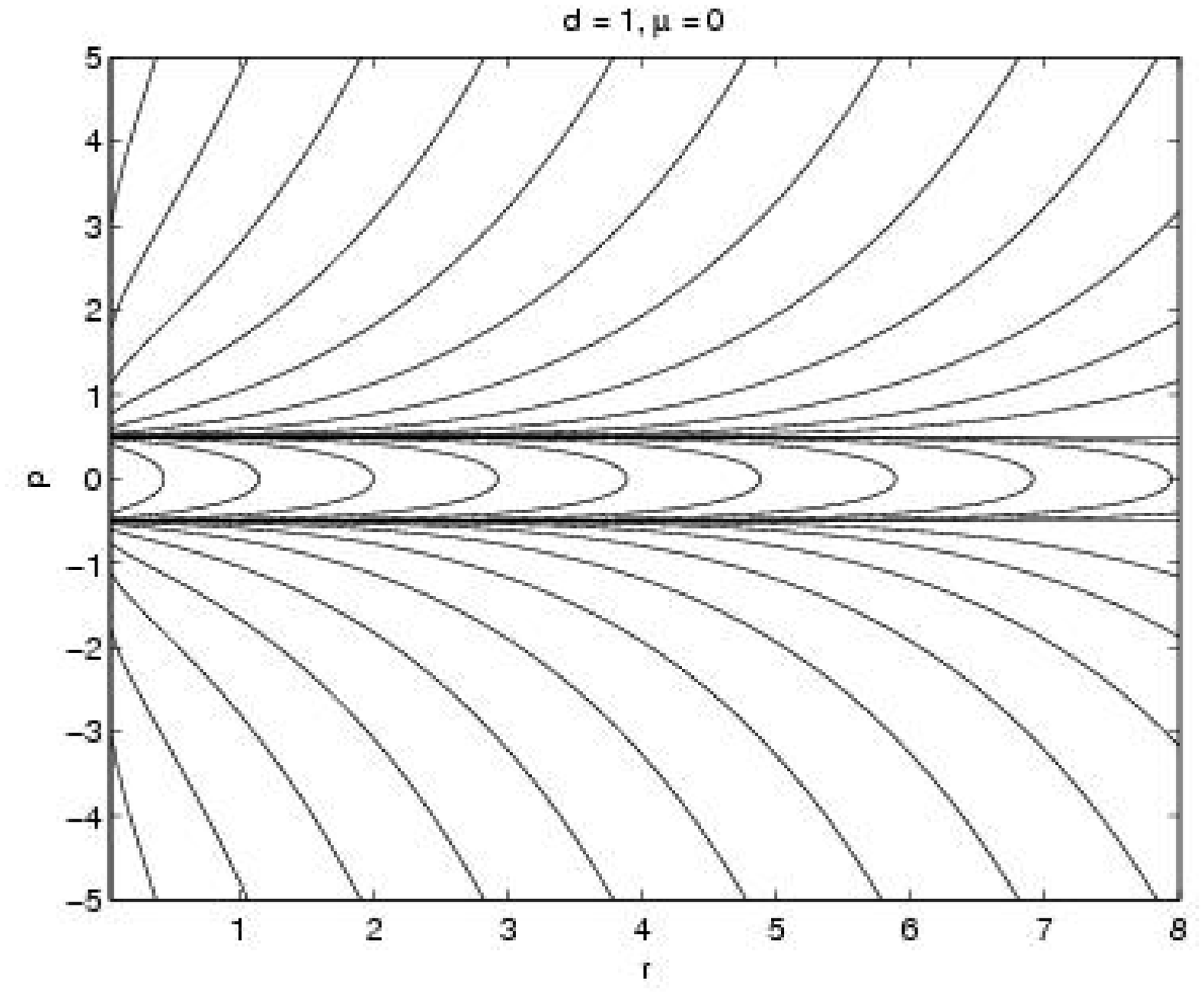}
\includegraphics[width=8cm]{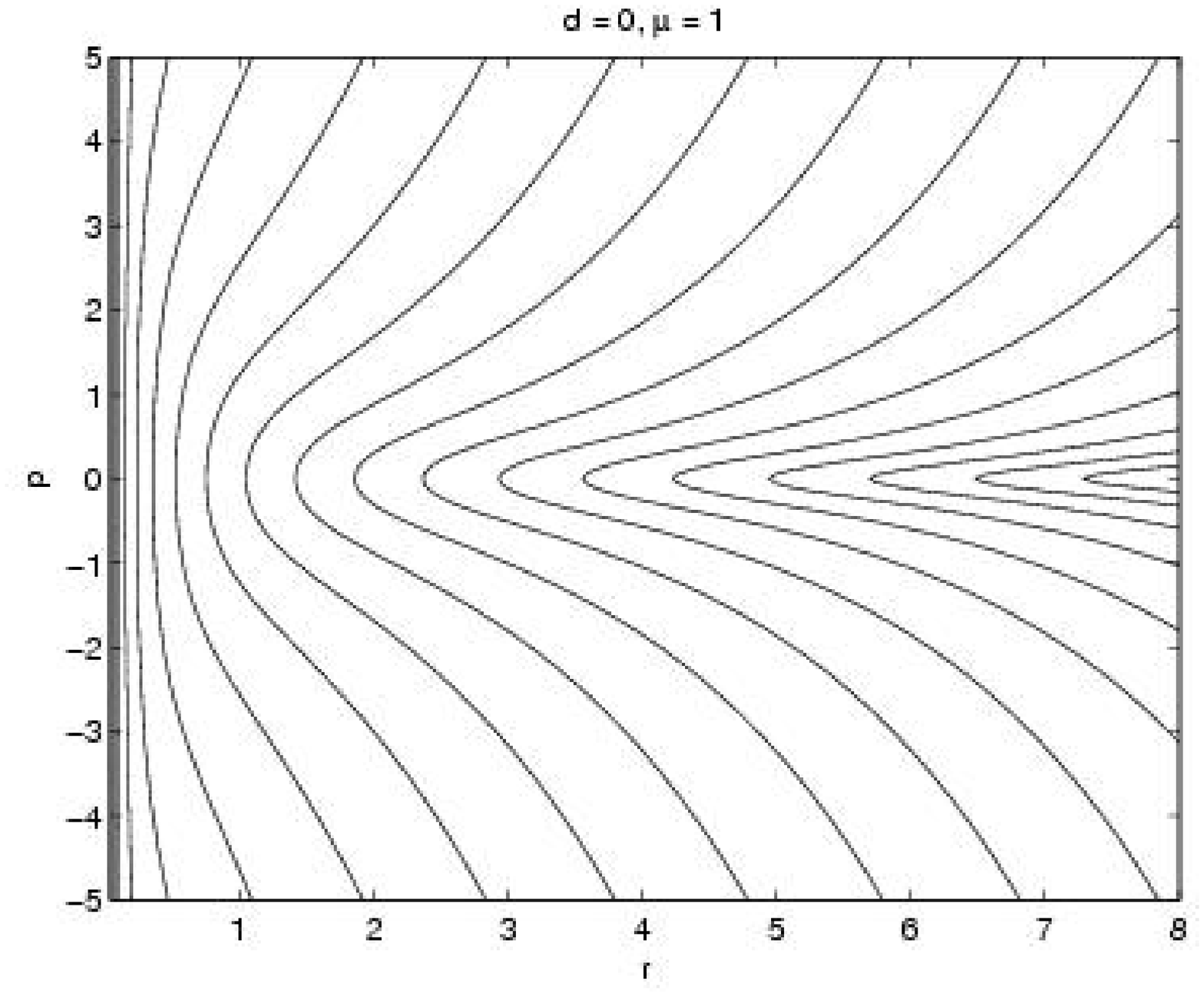}
\includegraphics[width=8cm]{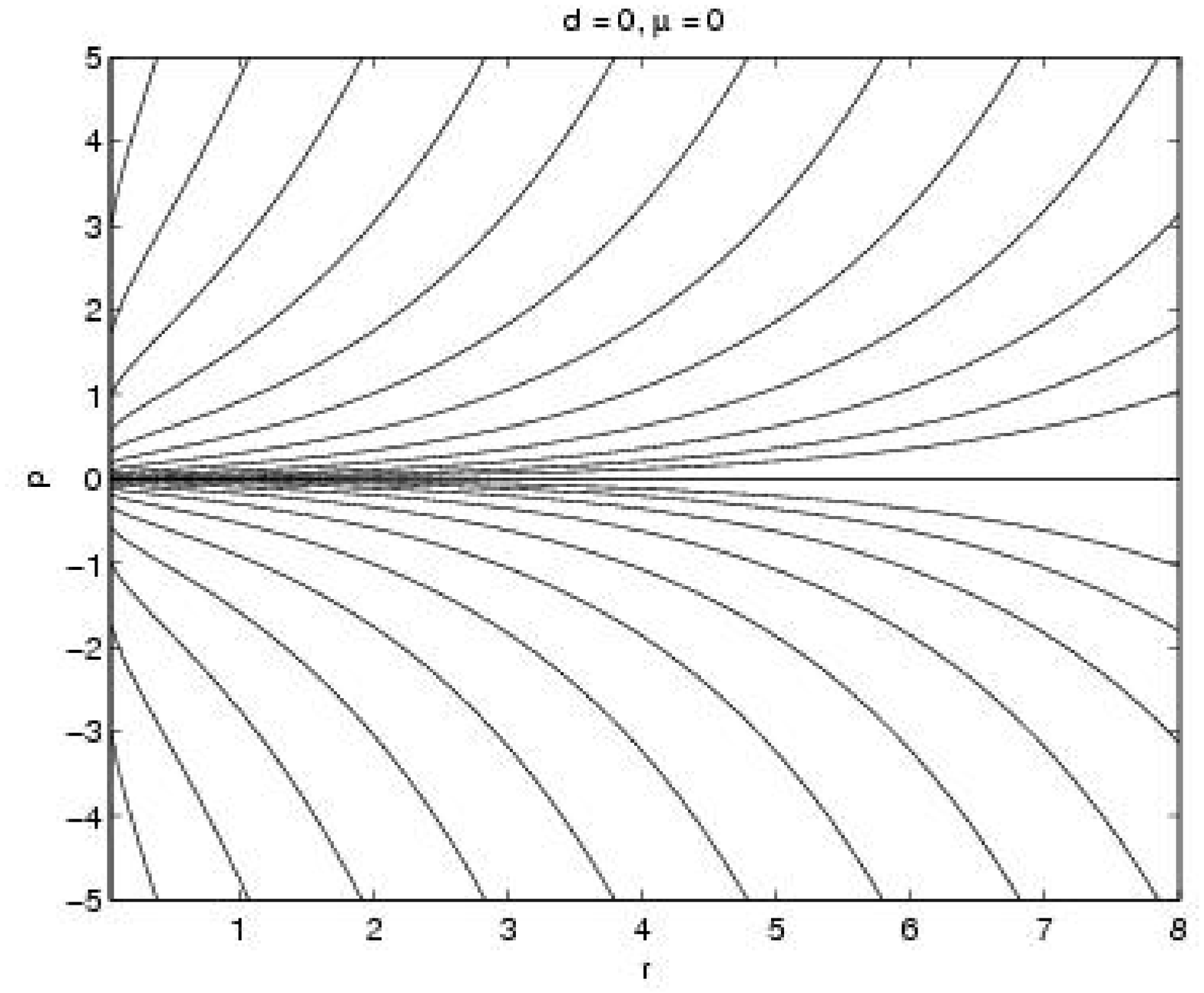}
\caption{Phase portrait for the 2 body problem with inertia operator ${\mathcal A}_1$ and regularization $G(0)=0$. The dynamics is somewhat reminiscent of traditional 2-body problems like the Kepler problem of Newtonian gravity: there are no capture orbits; for nonzero angular momentum (left) there are scattering and periodic orbits, while for zero angular momentum (right) genuine collisions are possible.}
\label{phase1}
\end{center}
\end{figure}

Setting $G(0)=0$ changes the character of the two body problem completely. It now features periodic orbits and scattering orbits, but no capturing orbits (see Figure~\ref{phase1}). However, numerical simulations of the {\it three} body problem in this case indicates that orbits can reach a singularity in a finite time, via a mechanism in which 2 particles orbit closer and closer, under regulation from the approach of the third particle.

\section{Capture orbits and dipoles\label{capture}}

Consider the $N$-body problem together with a smooth kernel so that 2-body captures are possible. Suppose that particles 1 and 2 become close enough to capture each other and that there are no other nearby particles. We will call the limiting state of such a captured pair of particles a `dipole'. In centre-of-mass coordinates for particles 1 and 2 only, the dipole is described by its centre of mass $c$, its momentum $d$, and its orientation $Q$ (equivalently by $(r,\theta)$) with conjugate momentum $P$ (equivalently $(p,p_\theta)$). From Eq. \ref{Happrox} we get that the Hamiltonian of a lone dipole is given by 
$$H_{1,2} := -\frac{1}{2}G''(0)(r^2p^2 + p_\theta^2) + \frac{1}{2}G(0)\|d\|^2.$$
The motion of a lone dipole is given by
\begin{equation}
\label{dipolemotion} 
c(t) = c_0 + t G(0) d_0,\quad d(t) = d_0,\quad
r(t) = r_0 e^{-\alpha t}, \quad p(t) = p_0 e^{\alpha t}, \quad \theta(t) = \theta_0 + \omega t,
\end{equation}
where the frequency $\omega=-G''(0)\mu$ and $\alpha = G''(0)p_0 r_0>0$ ($G''(0)<0$,
$p_0<0$).
However, other particles couple to all the degrees of freedom of the dipole so that it continues to undergo internal evolution.

Writing $H_{1,\dots,N}$ for the Hamiltonian of particles $1$ through $N$, the terms in $H_{1,\dots,N}$ coupling the dipole to particle $j$ are:
\begin{eqnarray}
H^d_j &:=& p_j\cdot \left[
\left(\frac{1}{2} d + P\right) G\left(\|c-q_j + \frac{1}{2}Q\| \right) + 
\left(\frac{1}{2} d - P\right) G\left(\|c-q_j - \frac{1}{2}Q\| \right) 
 \right] \nonumber \\
 &=& (d\cdot p_j) G(\|c-q_j\|) + (p_j\cdot P)((c-q_j)\cdot Q) \frac{G'(\|c - q_j\|)}{\|c-q_j\|}, \nonumber
 \end{eqnarray}
 plus terms of order $\mathcal{O}(e^{-\alpha t})$, which we omit if the dipole is taken to be in its limiting state. The first term in $H_j^d$ is the Hamiltonian for a single particle at position $c$ with momentum $d$, while the second (dipole) term generates an $\mathcal{O}(1)$ contribution to all of  $\dot d$, $\dot p_j $, $\dot q_j $,  $\dot Q/Q$, and $\dot P/P$. 
 Identifying $q_{N+1}\equiv c$ and $p_{N+1} \equiv d$, the Hamiltonian for a dipole
 and $N-2$ singleton particles is
\begin{eqnarray}
H_{1,\dots,N} &=& \frac{1}{2} \sum_{i,j=1}^N p_i\cdot p_j G(\|q_i - q_j\|) \nonumber \\
&=& \frac{1}{2}H_{1,2} + \frac{1}{2}H_{3,\dots,N+1} + \frac{1}{2}\sum_{j=3}^N H^d_j. \nonumber
\end{eqnarray}

This suggests that the overall evolution of $N$ particles with a smooth kernel is described by (i) free, straight-line motion of all particles when all particles are well-separated; (ii) scattering and interactions when distances are moderate; and (iii) capture or sticking together of 2 particles when they come within range, after which they continue to evolve as a dipole (and interact with the other particles). It is also possible for more than 2 particles to simultaneously capture each other. We have not analyzed this situation but it seems to follow the same course as the 2 body capture, with all inter-particle distances tending to 0 exponentially fast. It may also be possible for the influence of the other particles to cause a dipole to separate, although we never observed this happening. 

If the boundary conditions are taken to be periodic, this indicates that most orbits of the $N$-particle problem end up with all particles stuck together in one lump.

\begin{figure}
\begin{center}
\includegraphics[width=9cm]{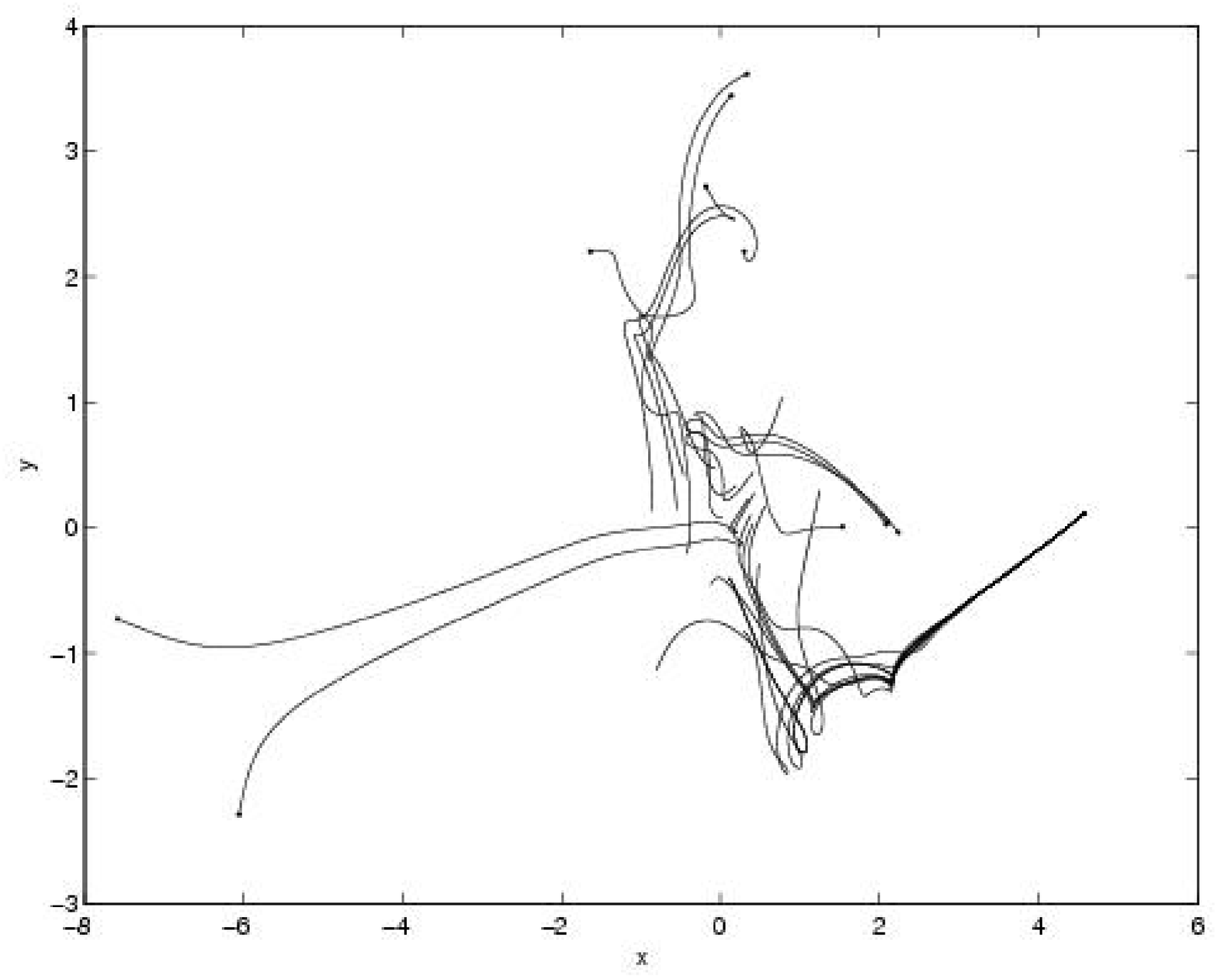}
\includegraphics[width=16cm]{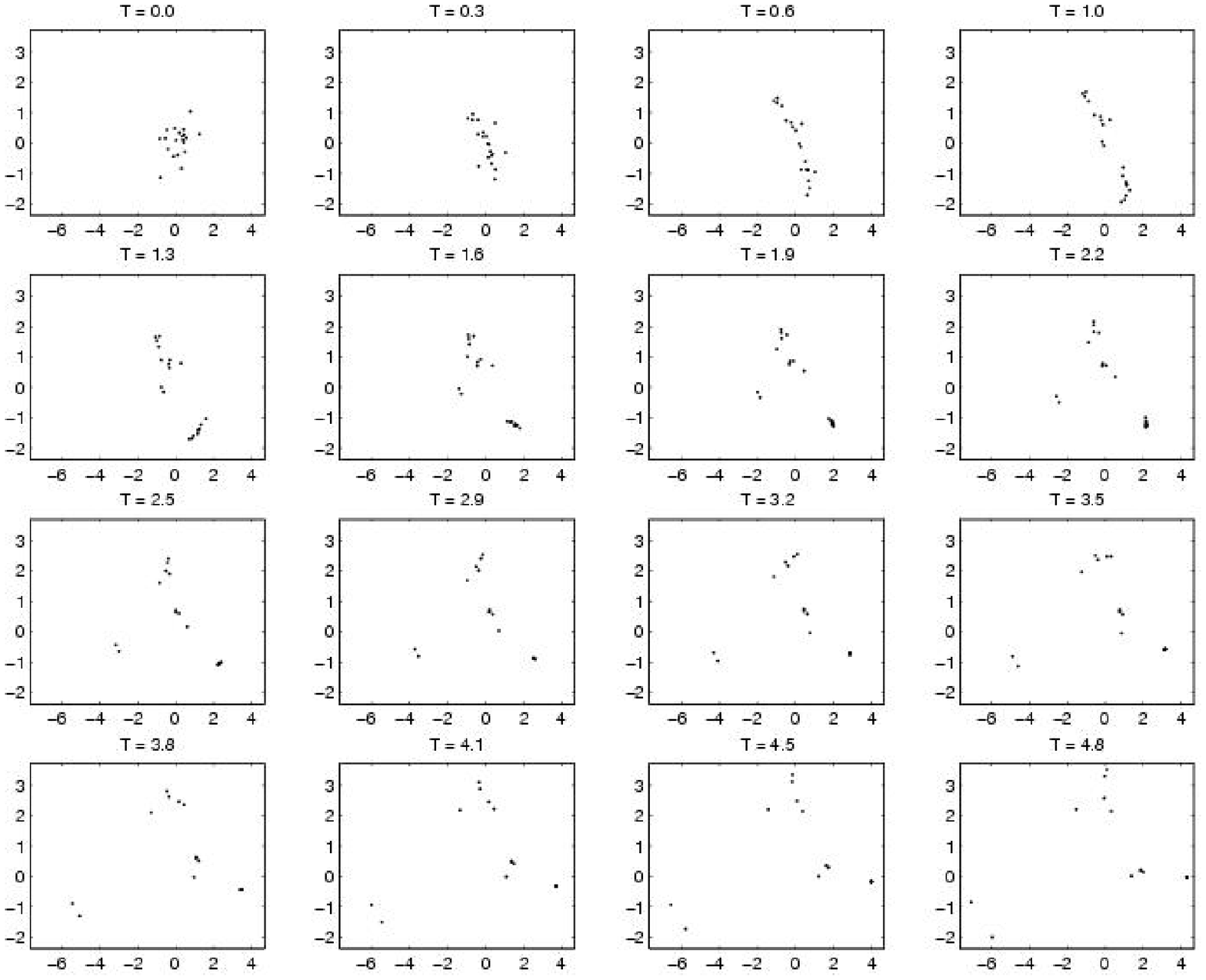}
\caption{Typical orbit for 20 randomly placed particles under a Gaussian kernel. (Top: particle paths, with final particle positions indicated by a dot; bottom: snapshots of the particle positions at 16 different times.)}
\label{orbit}
\end{center}
\end{figure}

A typical orbit for 20 particles in the plane is shown in Figure~\ref{orbit}. The initial positions and velocities are (normally) randomly distributed. Here the particles have been captured into groups of 9, 3 (with a 4th about to join), 2 (with 2 more possible captures), and 3 singletons.

We now analyze the influence of a dipole on a third particle with zero momentum---a
`test particle'. The test particle does not affect the motion of the dipole. We only have to
find the position $q$ of the test particle, which obeys
\begin{equation}
\label{eq:testpcle}
\dot q = d(t) G(\rho) + P((c(t)-q)\cdot Q) f(\rho),
\end{equation}
where $c(t)$ and $d(t)$ are given by Eq. (\ref{dipolemotion}),
$\rho = \|c-q\|$ is the distance from the dipole to the test particle, and
$f(\rho) = G'(\rho)/\rho$. 

One solution of (\ref{eq:testpcle}) is $q(t)=c(t)$, in which the test particle sits on top of the dipole. We shall see that dipoles can capture test particles into this state.

\begin{figure}
\begin{center}
\includegraphics[width=10cm]{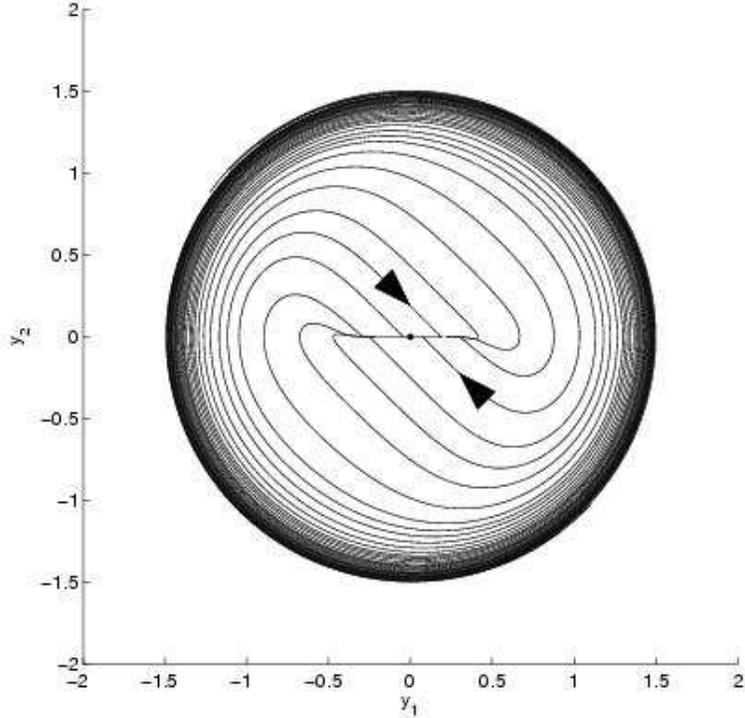}
\caption{Phase portrait of Eq. (\ref{eq:red2}) for a test particle in the field of a motionless dipole in rotating coordinates. Here, $\mu = 1$, $r_0p_0 = -1$, and $G(\rho) = e^{-\rho^2}$. \label{fig:test2}}
\end{center}
\end{figure}

First consider the case $d=0$ of a motionless dipole located at $c=0$. We compute
$$\eqalign{rP &= \left(\matrix{
p_0 r_0 \cos\omega t - \mu\sin\omega t \cr
p_0 r_0 \sin\omega t + \mu\cos\omega t }\right) \cr
&= A^{-1}\left(\matrix{-\sin\beta \cr \cos\beta }\right)\cr
Q &= \left(\matrix{ r \cos\omega t \cr r \sin \omega t}\right),
}$$
where $A^2 = p_0^2 r_0^2 + \mu^2$ and $\beta = \omega t + \phi$,
and the phase $\phi$ is defined by $\cos\phi = \mu/A$, $\sin\phi = -p_0r_0/A$.
Eq. (\ref{eq:testpcle}) becomes
$$ \dot q = -A^{-1} (q_1 \cos\omega t + q_2 \sin\omega t)f(\rho)
\left(\matrix{-\sin\beta \cr \cos\beta }\right);
$$ 
note the $e^{\alpha t}$ factors cancel out and the interaction is characterized
by the parameters $r_0p_0$ and $\mu$ of the dipole.
The form of these equations suggests going into a rotating frame by
$$ y = \left(\matrix{ \cos\beta & \sin\beta \cr -\sin\beta & \cos\beta}\right)
q,$$
which leads to a drastic simplification to the planar, autonomous system
\begin{equation}
\label{eq:red2} \eqalign{\dot y_1 &= \omega y_2, \cr
\dot y_2 &= -(\mu y_1 + r_0p_0 y_2)f(\rho) - \omega y_1. \cr
}\end{equation}
Note $\rho = \|q\| = \|y\|$. The phase portrait of this system is shown in Figure \ref{fig:test2}.

We now investigate the stability of the fixed point $y=0$ under the assumption that $G$ is smooth, so that $f(\rho) = G''(0) + \frac{1}{6}G^{(4)}(0)\rho^2 + \dots$ and (\ref{eq:red2}) becomes
$$ \eqalign{\dot y_1 &= \omega y_2= -G''(0)\mu y_2 \cr
\dot y_2 &= r_0 p_0 G''(0)y_2 - \frac{1}{6}G^{(4)}(0)(y_1^2+y_2^2) + o(\|y\|^2).\cr
}$$
The eigenvalues of the fixed point $y=0$ are 0 (eigenvector $(1,0)^T$) and
$-p_0r_0G''(0)<0$ (eigenvector $(\mu,p_0r_0)^T$). The fixed point is nonhyperbolic, but a routine application of centre manifold theory
then gives a centre manifold located at $y_2 = \gamma y_1^3 + \mathcal{O}(y_1^4)$ with
$ \gamma = - \mu G^{(4)}(0)/(6 p_0 r_0 G''(0))$, and, on the centre manifold, the dynamics
is governed by the reduced equation
$$ \dot y_1 = \delta y_1^3,\quad \delta = \omega\gamma = \frac{\mu^2 G^{(4)}(0)}{6 p_0 r_0}.$$
Recalling that $p_0<0$ for a dipole, we have that $y=0$ is asymptotically stable
if $G^{(4)}(0)>0$ and unstable if $G^{(4)}(0)<0$. The fixed point is stable for the
Gaussian kernel $G(\rho) = e^{-\rho^2}$.

\begin{figure}
\begin{center}
\includegraphics[width=10cm]{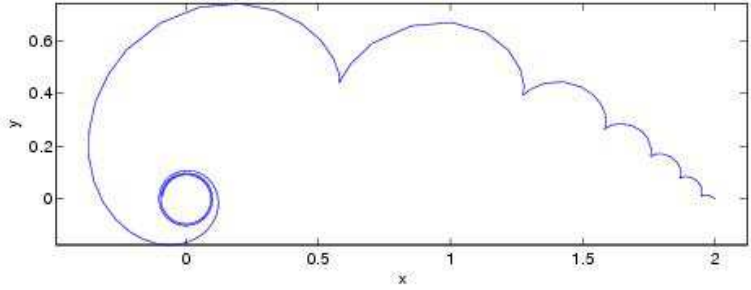}
\caption{Attraction of a test particle to a dipole at the origin. (The particle moves (under Eq. (\ref{dipolemotion}) in the direction of the rotating direction vector of the dipole, with the cusps corresponding to the dipole pointing directly at the particle.)\label{fig:test1}}
\end{center}
\end{figure}

That is, the dipole for a Gaussian kernel attracts nearby test particles, albeit very slowly,
since $y_1(t) = (y_1(0)^{-2} - 2 \delta t)^{-1/2}$. A typical test particle far from the dipole moves initially slowly (since the velocity field is exponentially small there), then quickly falls onto the centre manifold, then orbits the dipole at the same frequency as the dipole itself, maintaining a phase shift of $\phi$, at a gradually diminishing distance. See Figures \ref{fig:test1} and \ref{fig:test2}.

\begin{figure}
\begin{center}
\includegraphics[width=16cm]{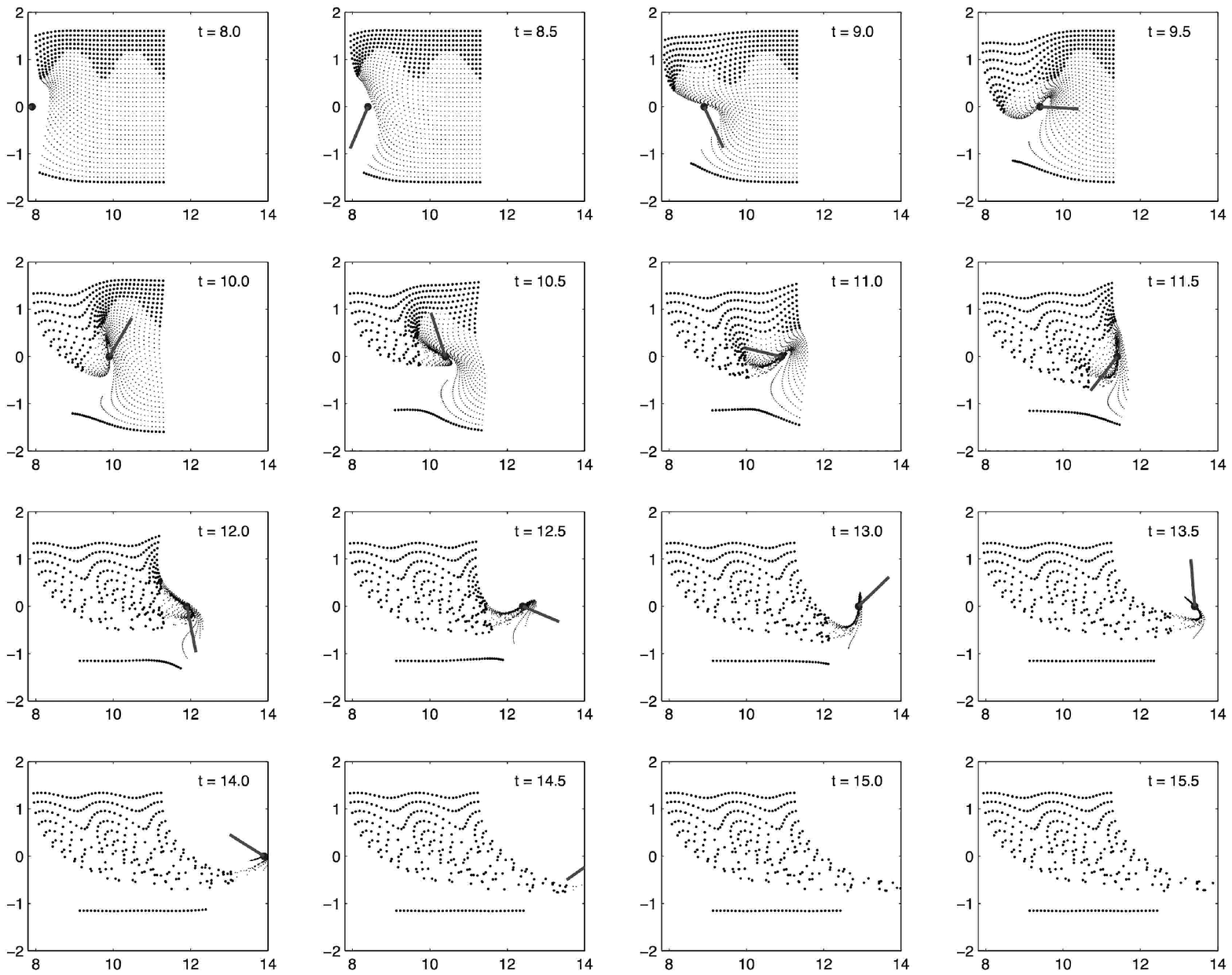}
\caption{Entrainment of a field of test particles by a moving dipole. The dipole (whose direction is indicated by a stick) is moving to the right with speed 1.  Particles shown as small dots are entrained by the dipole, while particles shown as larger dots are left behind.\label{fig:ent}}
\end{center}
\end{figure}

Finally, we consider a moving dipole, which we take (e.g., by rotating space and by scaling time) to be moving along the $x$-axis at unit speed, $c(t) = t(1,0)$. A test particle with sufficiently large $|q_2(0)|$ (i.e., with the test particle far enough from the path of the dipole) will not be entrained as the dipole passes. However, the critical value of $q_2(0)$ will also depend on $q_1(0)$, which determines the phase of the dipole as it approaches. A numerical simulation (see Figure \ref{fig:ent}) reveals a surprising asymmetry where, for an anticlockwise ($\mu>0$) dipole, the entrainment is independent of the phase for $q_2(0)<0$ and strongly dependent on the phase for $q_2(0)>0$. All particles out to $q_2 = -1.536$, on the right of the dipole, are entrained, but on the left the critical initial distance varies from $0.6$ to $1.15$.

\section{Discussion}

The particle dynamics in $\diff(\R^2)$ with a $H^k$ metric ($k \ge 2$) that we have considered in this paper show striking contrasts to the more familiar dynamics of point vortices for $\diff_{\rm vol}(\R^2)$ with an $L^2$ metric. Of particular note are:
\begin{itemize}
\item our particles  are vector particles, point vortices are point particles;
\item for our particles, the 2 body problem is integrable, for point vortices the 3 body problem is integrable;
\item our particles exhibit short-range interactions $u(r)\sim e^{-r} r^{k-1/2}$ ($r\to\infty$), point vortices exhibit long-range interactions $u(r)\sim 1/r$;
\item our particles have no periodic 2-body orbits, whereas most 2-body orbits are periodic for point vortices; and
\item the 2-body dynamics of $\diff(\R^2)$ depend sensitively on the metric, whereas for point vortices, the {\em qualitative} 2-body dynamics are independent of the metric.
\end{itemize}

Clearly, we have only scratched the surface of the rich dynamics of the system (\ref{eq:eom}). Further dynamical questions to be considered include
\begin{itemize}
\item a full study of the 3 body problem;
\item determination of the long-time existence of solutions;
\item a classification of the limiting states of the $N$-body problem;
\item a study of the limiting (`$N$-pole') state of the capture of $N$ particles; and
\item a consideration of the geometry of the Riemannian manifold whose geodesics are governed by (\ref{eq:eom}),
\end{itemize}
while broader questions include
\begin{itemize}
\item the significance of the sensitivity to the metric for image matching applications; and
\item the convergence of the point particle approximation to solutions of the PDE
(\ref{Euler1},\ref{Euler2}).
\end{itemize}

\bibliographystyle{plain}
\bibliography{Euler}

 \end{document}